\documentclass[aps,reprint,groupedaddress,prb]{revtex4-2}

\usepackage{graphicx}
\usepackage[colorlinks,bookmarks=false,linkcolor=red,urlcolor=red, citecolor=green]{hyperref}
\usepackage{amsmath}
\usepackage{amsfonts}
\usepackage{dsfont}

\newcommand{\bra}[1]{\langle#1|}
\newcommand{\ket}[1]{|#1\rangle}
\newcommand{\braket}[2]{\langle#1|#2\rangle}

\newcommand{\real}[1]{\mathrm{Re}\left[#1\right]}

\newcommand{\expval}[2]{\mathbb{E}_{#2}\left[#1\right]}
\newcommand{\tr}[0]{\mathrm{Tr}}

\newcommand{\qexp}[1]{\langle#1\rangle}
\newcommand{\psit}[0]{\psi_{\boldsymbol\theta}}

\usepackage{color}

\definecolor{nqdcolor}{rgb}{0.5586, 0.0586, 0.4219}
\definecolor{blue}{rgb}{0.384, 0.553, 0.941}

\usepackage{comment}

\begin{document}

\title{Modeling light-matter coupled systems with neural quantum states}

\author{Noe Salmeron$^1$}
\email{noe.salmeron@ur.de}
\author{Marin Bukov$^2$}
\thanks{equal contribution}
\author{Markus Schmitt$^{1,3}$}
\thanks{equal contribution}

\affiliation{$^1$University of Regensburg, Universitätsstr. 31, Regensburg D-93053, Germany}
\affiliation{$^2$Max Planck Institute for the Physics of Complex Systems, Nöthnitzer Str. 38, 01187 Dresden, Germany}
\affiliation{$^3$Institute of Quantum Control (PGI-8), Forschungszentrum Jülich, D-52425 Jülich, Germany}

\date{\today}

\begin{abstract}

Recent advances in cold atom manipulation enable the study of many-body systems where short-range interactions between neighboring atoms coexist with long-range interactions mediated by photons. Such a combination of interactions makes a theoretical approach challenging beyond mean-field methods. In this work, we develop a neural quantum state based approach to study these systems numerically. We introduce a neural-network architecture capable of handling hybrid Hilbert spaces with large local bosonic dimensions in strongly interacting spin–photon systems. We benchmark this approach on a model of a two-dimensional lattice of Rydberg atoms coupled to a photon mode. The superradiant ground states found in the large spin-photon coupling regime allow us to demonstrate the efficiency of the method in the presence of high photon occupation. Furthermore, the ability to capture spin-spin and spin-photon correlations leads us to observe quantitative deviations in the ground state phase boundaries with respect to mean-field theory. The method extends to other systems with a similar hybrid Hilbert space structure, such as spin-phonon systems, and provides a scalable framework for investigating their ground state properties.
\end{abstract}

\maketitle

\section{Introduction}

\begin{figure*}[]
\includegraphics[width=1\textwidth]{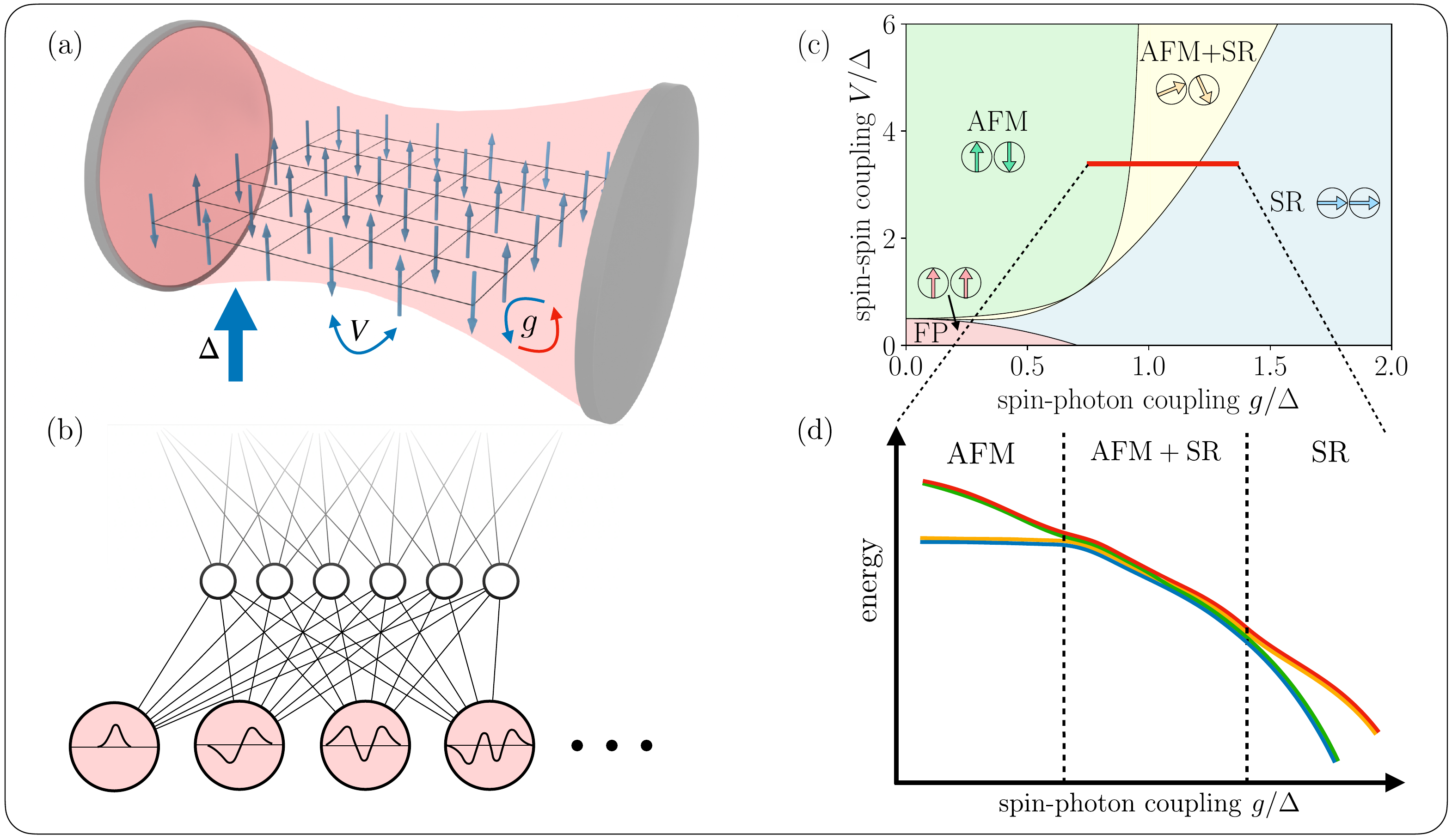}
\caption{\label{fig:fig1} (a) 2D array of Rydberg atoms, represented as two-level systems, in a cavity. The atoms on the lattice interact repulsively with their nearest neighbors with strength $V$. They are additionally uniformly coupled to a single cavity photon mode with interaction strength $g$. The cavity effectively mediates an all-to-all coupling competing with the nearest-neighbor interaction. An external field with intensity $\Delta$ is also present. For electrically neutral Rydberg atoms, this field can take the form of an external periodic drive~\cite{schauss2012observation}. (b) We study the ground state properties using Neural Quantum States (NQS) by designing a neural network architecture capable of handling both spin and photon degrees of freedom.
(c) Phase diagram predicted by mean-field theory for a cavity frequency $\omega_0/\Delta=2.0$. It features a fully polarized along $z$ (FP), an antiferromagnetic (AFM), and a superradiant (SR) phase. It also predicts a mixed phase (AFM+SR) that lies between the AFM and SR phases and contains properties of both phases. It is separated into two regions that are joined by a single multi-critical point at $V/\Delta=1$. The arrows illustrate the mean-field spins orientation on the even and odd sublattices. See Appendix~\ref{app:MF coherent} for the full mean-field derivation. (d) The symmetries of the model allow us to sketch the energy level structure and predict the ground state degeneracy. The AFM phase breaks the unit cell translation symmetry $\mathcal T$ so the ground state manifold hosts a symmetric (in blue) and an antisymmetric (in orange) state. The SR phase, on the other hand, breaks the symmetry $\mathcal P: (a, \sigma^\pm)\mapsto(-a, -\sigma^\pm)$, giving rise to a symmetric (in blue) and anti-symmetric (in green) ground states. The mixed AFM+SR phase breaks the two symmetries simultaneously, making the ground state four-fold degenerate.
}
\end{figure*}

In recent years, important progress has been made in the ability to control and manipulate neutral atoms, enabling the realization of increasingly sophisticated quantum simulators~\cite{browaeys2020many, bloch2012quantum}. These platforms promise to offer insights into a broad range of complex many-body quantum phenomena. A particularly active line of research involves embedding these simulators into optical cavities, where local interactions between atoms coexist with collective interactions mediated by cavity photons~\cite{landig2016quantum, klinder2015observation, marsh2025multimode}. The competition between short-range and long-range interactions is expected to give rise to a rich variety of quantum phases of matter and collective phenomena~\cite{defenu2023long}. At the same time, this interplay makes such systems difficult to study analytically.

Efficient numerical methods are therefore important for capturing effects beyond mean-field, but not all are applicable. Exact diagonalization is limited to small system sizes. Tensor network methods face difficulties in two dimensions and with long-range or all-to-all interactions. Quantum Monte Carlo methods are applicable in some cases, but their use is restricted when a sign problem arises. Neural quantum states (NQS) have recently emerged as a powerful variational framework; they have been extensively applied to two-dimensional strongly correlated lattice models, where other conventional methods encounter difficulties~\cite{carleo2017solving, white1992density, deng2017quantum}. Little work has been done with NQS, however, for systems involving degrees of freedom of different nature, such as light-matter systems~\cite{mahajan2025structure, lagnese2026neural}. The presence of strong interactions between degrees of freedom, some having an infinite local Hilbert space dimension, poses major challenges for constructing adequate and efficient neural-network ans\"atze.

Beyond light-matter systems, similar hybrid structures arise in a variety of physical systems. Examples include phonons interacting with matter, such as in trapped ion platforms, or even lattice gauge theories with truncated gauge-field Hilbert spaces~\cite{Chandrasekharan1997}.

In this work, we introduce an NQS-based approach designed to address such hybrid systems. The neural network is tailored to represent an extensive number of strongly correlated two-level systems coupled to a bosonic mode by introducing a multi-head neural network architecture capable of handling both the spin and the bosons degrees of freedom. This allows for efficient variational simulations of systems beyond the reach of exact diagonalization, even for large occupation numbers in the bosonic sector. More generally, this approach provides a systematic framework for extending NQS to a wide class of hybrid quantum systems.

As a concrete application, we study a two-dimensional Rydberg array coupled to a cavity photon mode~\cite{gelhausen2016quantum}. Its ground state can host a large number of photons, with an occupation that increases with system size, thereby providing a challenging test for our method. This system, also known as the Dicke-Ising model, has generated substantial interest. Theoretical approaches have enabled the derivation of ground-state properties in the low-photon-occupation limit~\cite{schellenberger2024almost, bezvershenko2021dicke}, while numerical approaches, including ours, enable predictions beyond this limit~\cite{an2022quantum, langheld2025quantum}. This model has been realized experimentally in one dimension, and efforts to realize it in two dimensions are underway~\cite{de2026realization, zhang2014quantum}.

Using the proposed NQS approach, we capture correlations beyond mean-field and investigate the resulting ground-state properties. Our results show quantitative deviations in the location of the phase transitions predicted by mean-field theory, highlighting the important role of correlations in this regime. We demonstrate that neural quantum states provide a viable and flexible tool for studying two-dimensional light–matter systems, even in regimes with large photon occupation. The method can then naturally be generalized to other systems with similar hybrid Hilbert spaces.

This paper is organized as follows. 
Section~\ref{sec:model} introduces the physical model and reviews its mean-field solution, which we use as a reference for our numerical results. Section~\ref{sec:methods} briefly introduces the neural quantum states framework and showcases the neural network architecture we use. This is followed by a presentation of several techniques designed to improve the training procedure, as well as a discussion regarding the choice of the photon number cutoff, an important hyperparameter. Finally, the results are presented and discussed in Section~\ref{sec:results}.

\section{Dicke-Ising model of a Rydberg atom array in a cavity}
\label{sec:model}

The model we consider is inspired by recent proposals to investigate two-dimensional arrays of Rydberg atoms coupled to a single mode of an optical cavity~\cite{gelhausen2016quantum}. It is relevant for our purposes because it hosts a superradiant regime characterized by a large photon occupation in the ground state. In this regime, the photon number grows with system size, making numerical simulations especially challenging and providing a stringent test for our method.

Notice that accurate steady state descriptions of the system would require accounting for dissipative effects such as photon loss in the cavity \cite{gelhausen2016quantum,bezvershenko2021dicke}. The Hamiltonian description used in the following instead targets intermediate time windows upon an adiabatic preparation protocol to realize low-energy states. This time window may, in fact, be rather long, as the approach towards the dissipative steady state is rather slow for large systems \cite{bezvershenko2021dicke}.

\subsection{Hamiltonian}
\label{sec:hamiltonian}

The $N$ Rydberg atoms are described as two-level systems interacting with their nearest neighbors, leading to the following Ising-like Hamiltonian for the spins:
\begin{equation*}
    H_{\text{spin}} = -\frac{\Delta}{2} \sum_{\ell=1}^{N} \sigma_{\ell}^{z} + \frac{V}{d} \sum_{\langle \ell m \rangle} \left( \frac{1 + \sigma_{\ell}^{z}}{2} \right) \left( \frac{1 + \sigma_{m}^{z}}{2} \right),
\end{equation*}
where $[\sigma^\alpha_i,\sigma^\beta_j]=2i\varepsilon^{\alpha\beta\gamma}\sigma^\gamma_i\delta_{ij}$ are the Pauli matrices.
The first term describes an external field of strength $\Delta$, which we use to set the energy unit throughout. The second term represents a sum over nearest-neighbor interactions, denoted by $\langle \ell m\rangle$. It uses shifted $\sigma^z$ operators to indicate that two neighboring Rydberg atoms in the blockade regime interact only when both are excited. Rydberg atom arrays typically exhibit longer-ranged van-der-Waals interactions, but we consider the simplified model restricted to nearest-neighbor interactions. We will only consider positive values of $V$ for which the interaction favors an antiferromagnetic ordering.

The spin system is placed in a cavity and interacts with a single mode of the electromagnetic field, described by the creation and annihilation operators $[a,a^\dagger]=1$, and the photon number operator $n=a^\dagger a$. The cavity Hamiltonian is 
$$H_{\text{cavity}} = \omega_0\; a^\dagger a,$$ 
and the homogeneous spin-photon interaction term with coupling $g$ is
\begin{equation*}
    H_{\text{int}} = \frac{g}{\sqrt{N}} (a + a^{\dagger}) \sum_{\ell=1}^{N} (\sigma_{\ell}^{+} + \sigma_{\ell}^{-}).
\end{equation*}
The $1/\sqrt N$ scaling of the coupling strength ensures a non-trivial thermodynamic limit by keeping the interaction energy extensive.
The Hamiltonian of the entire system reads as 
\begin{equation}
\label{eq:full H}
    H = H_{\text{spin}}+H_{\text{cavity}}+H_{\text{int}}.
\end{equation}
A detailed description of how each term can be implemented experimentally can be found in Ref~\cite{gelhausen2016quantum}.

When expressed in the basis $\{\ket{n,\sigma_1,\sigma_2,\dots,\sigma_N}\}$ with $n=0,1,2,\dots$ and $\sigma_j=\{1,-1\}$ the eigenvalue of the Pauli operator $\sigma^z_j$, non-zero off-diagonal elements of the Hamiltonian appear only in $H_{\text{int}}$. 
In this work, we focus on positive values of the spin-photon coupling $g$, so these non-zero off-diagonal elements are positive, and the Hamiltonian is non-stoquastic~\cite{bravyi2014monte}. This is known to present difficulties for numerical simulations~\cite{troyer2005computational, bravyi2014monte}. To remedy this, we apply the following unitary transformation to obtain a stoquastic Hamiltonian:
\begin{equation}
\label{eq:U_a}
    U_a = \sum_j(-1)^j \ket{j}\bra{j}\otimes\mathds{1}_{\text{spins}}.
\end{equation}
It acts on the photon degree of freedom only and satisfies $U_a a^\dagger U_a^\dagger = -a^\dagger$ and $U_a a U_a^\dagger = -a$, effectively flipping the sign of the interaction Hamiltonian $H_\text{int}$. The stoquasticity of $\tilde H=U_aHU_a^\dagger$ ensures that the ground state wavefunction can be chosen real and positive~\cite{bravyi2014monte}. In the context of the NQS approach, it allows us to disregard the phase of each wavefunction coefficient, and to only represent their amplitude, as detailed in Section~\ref{sec:architecture}.

In this work, we therefore focus on finding the ground state $\ket{\tilde \psi}$ of the transformed Hamiltonian $\tilde H$. The ground state $\ket{\psi}$ of the original Hamiltonian is recovered via $\ket{\psi} = U_a \ket{\tilde \psi}$. Expectation values of observables are computed as $\bra{\tilde \psi} U_a \mathcal{O} U_a^\dagger \ket{\tilde \psi}$, which simply reduces to $\bra{\tilde \psi} \mathcal{O} \ket{\tilde \psi}$ for all spin observables and for $\mathcal O=a^\dagger a$. Furthermore, since $U_a^2=\mathds 1$, the reverse transformation is simply $U_a^{-1}=U_a$.

The Hamiltonian has two $\mathbb{Z}_2$ symmetries that will be important throughout this work. 
First, because we consider periodic boundary conditions, $H$ is invariant under the unit cell translation symmetry $\mathcal T$, defined as translation by one lattice site along the horizontal or vertical direction.
The second symmetry, $\mathcal P: (a, \sigma^\pm)\mapsto(-a, -\sigma^\pm)$, is generated by the unitary
\begin{equation}
\label{eq:U_sr}
    U_{\mathcal{P}} = \left(\bigotimes_{\ell=1}^N \sigma_\ell^z\right)\otimes U_a
\end{equation}
with $U_a$ defined in Eq.~\eqref{eq:U_a} and $\bigotimes_{\ell=1}^N \sigma_\ell^z$ representing a $\sigma^z$ operator acting on each atom. This unitary also obeys $U_{\mathcal{P}}^2=\mathds{1}$ and therefore defines two symmetry sectors with eigenvalues $\pm 1$. We will later use both symmetries to compare ground state properties in different symmetry sectors.

\subsection{Mean-field solution}

With the Hamiltonian now defined, we briefly recap the ground state properties predicted by mean-field theory. This will allow us to gain intuition about the model and will serve as a basis for comparison with the NQS approach later. The details of the  mean-field derivation are given in Appendix~\ref{app:MF coherent}; for the case of a lossy cavity, we refer the reader to Refs.~\cite{gelhausen2016quantum,bezvershenko2021dicke}.

Figure~\ref{fig:fig1}(c) shows a slice of the mean-field phase diagram for a fixed value $\omega_0/\Delta=2$ of the cavity frequency~\cite{gelhausen2016quantum}. It shows four phases: 
(i) for low spin-spin coupling $V/\Delta$ and low spin-photon coupling $g/\Delta$, the external field term with strength $\Delta$ dominates, and the system is in the fully polarized (FP) phase. Then, (ii), keeping the spin-photon coupling small but increasing the spin-spin coupling leads to the antiferromagnetic (AFM) phase. To identify this phase, we use the staggered magnetization as an order parameter: $M_s\equiv\qexp{\sigma_{\text{even}}^z}-\qexp{\sigma_{\text{odd}}^z}$, where $\qexp{\sigma_{\text{even/odd}}^z}$ denotes the total magnetization on the even/odd sublattices, respectively. Next, (iii), large values of the spin-photon coupling give rise to the superradiant (SR) phase. The system is SR if, in the ground state, $\qexp{a}/\sqrt N\neq0$. Finally, (iv), mean-field theory predicts a mixed phase in which both AFM and SR order coexist. It is present whenever the order parameters of the AFM and of the SR phase are non-zero simultaneously. This coexistence is made possible by the composite nature of the system, which allows spin ordering even in the presence of photons.

The AFM phase breaks the unit cell translation symmetry $\mathcal T$ while the SR phase breaks the $\mathcal P$ symmetry defined in Eq.~\eqref{eq:U_sr}. Each of these $\mathbb{Z}_2$ symmetries, when broken individually, leads to a twofold degeneracy of the ground state. In the mixed AFM+SR phase, where both symmetries are broken simultaneously, the degeneracy is therefore fourfold. From these considerations, we show in Fig.~\ref{fig:fig1}(d) the expected energy level structure in the ground state in the thermodynamic limit. The four states correspond to the different symmetry sectors associated with the translation symmetry $\mathcal T$ and the parity symmetry $\mathcal P$.

Two asymptotic cases are of particular interest to gain more intuition, since they relate to extensively studied models: $g/\Delta=0$ gives the classical antiferromagnetic Ising model in an external field, and $V/\Delta=0$ corresponds to the Dicke model~\cite{dicke1954coherence}.

As shown in Fig.~\ref{fig:fig1}(c), the mean-field results predict that the mixed phase splits into two regions separated by a single multi-critical point at $V/\Delta=1$. The ground states predicted by mean-field theory, however, are limited to product states. One of our goals, therefore, is to examine how accounting for spin-photon and spin-spin correlations affects this picture of the system's behavior.

\section{Neural Quantum States for spin-photon systems}
\label{sec:methods}

Having introduced the model and its mean-field description, we now describe the variational framework we use to approximate its ground state. We combine variational Monte Carlo with an NQS ansatz tailored to the composite spin–photon Hilbert space.

\begin{figure*}[t]
\includegraphics[width=0.95\textwidth]{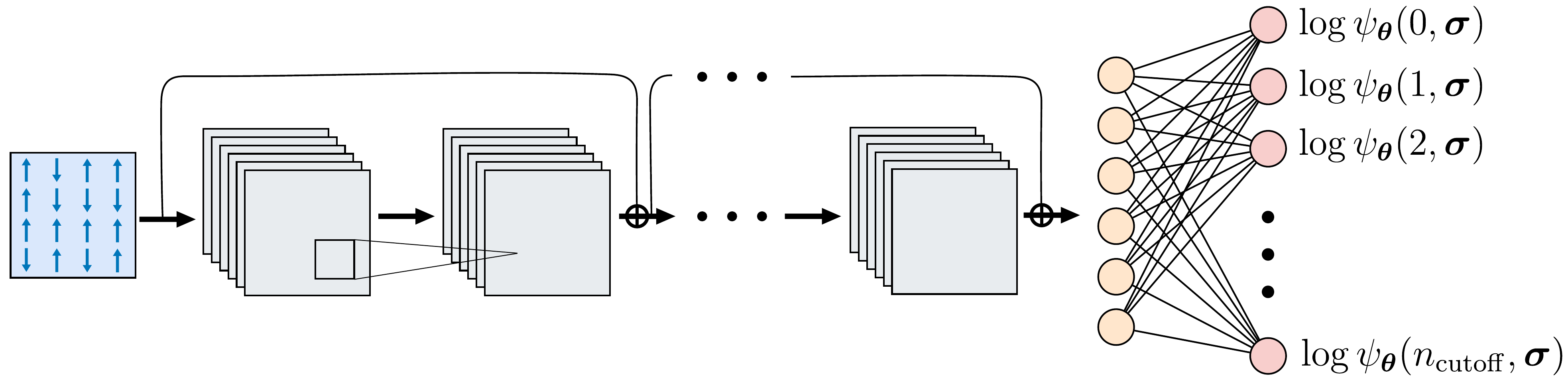}
\caption{\label{fig:architecture} Schematic illustration of the neural network architecture. We adapt the ResNet architecture previously used for spin-only systems~\cite{chen2024empowering}. A spin configuration is given as input and passes through multiple convolutional layers with residual connections. Our modification occurs at the end of these layers: instead of averaging all feature maps to a single number, we average each one separately to produce an array. These values are then fed into a final dense layer that outputs the log-wavefunction coefficients for each photon number in a multi-head fashion.
}
\end{figure*}

\subsection{Neural Quantum States}
\label{sec:nqs}

In this work, we use the $\{\ket{x}\}=\{\ket{n,\sigma_1,\sigma_2,\dots,\sigma_N}\}$ basis introduced in Section~\ref{sec:hamiltonian} as the computational basis. As we shall see, however, the variational Monte Carlo formalism we use can be formulated in a generic basis.

Exactly representing the full wavefunction of large many-body quantum states is impossible due to the exponential scaling of the Hilbert space with the number of degrees of freedom. The presence of bosonic modes worsens the problem by making the Hilbert space dimension infinite for any lattice size. The NQS approach remedies this by providing a variational representation of the wavefunction $\psi_{\boldsymbol\theta}(x)$ using an artificial neural network with parameters $\boldsymbol\theta$. Because neural networks are universal function approximators~\cite{hornik1989multilayer, lin2018resnet}, increasing their size allows them to represent the ground state arbitrarily precisely, at the cost of a larger number of parameters.

To make the Hilbert space finite, we truncate the infinite photon degree of freedom with a cutoff $n_{\text{cutoff}}$. In the regimes of interest, such as the superradiant phase, large cutoffs are needed to properly represent the ground state. Addressing this challenge by designing an architecture capable of dealing with an extensive number of photons is a key part of this work. We discuss extensively in Secs.~\ref{sec:architecture}~and~\ref{sec:cutoff} how this cutoff is implemented, how to choose it, and the associated challenges. 

\subsection{Variational Monte Carlo}
\label{sec:sr}

Given a variational state $\ket{\psi_{\boldsymbol\theta}}$, an artificial neural network in our case, its energy expectation value can be estimated using conventional variational Monte Carlo techniques. For this purpose, the expectation value is rewritten as
\begin{equation}
    \frac{\bra{\psit}H\ket{\psit}}{\braket{\psit}{\psit}}=\sum_{\{x\}}\frac{|\psit(x)|^2}{\braket{\psit}{\psit}}E_\text{loc}(x)=\expval{E_\text{loc}(x)}{\pi},
\end{equation}
where $\pi(x)=|\psit(x)|^2/\braket{\psit}{\psit}$ is the Born distribution, and the local energy is defined as:
\begin{equation}
\label{eq:Eloc}
    E_\text{loc}(x)=\sum_{\{x'\}}\bra{x}H\ket{x'}\frac{\psit(x')}{\psit(x)}.
\end{equation}
The expectation value can be estimated by Monte Carlo sampling from the Born distribution. This procedure remains tractable as long as the number of non-zero terms in the sum of Eq.~\eqref{eq:Eloc} scales polynomially with the system size. Typical Hamiltonians like the one defined in Eq.~\eqref{eq:full H} are sparse and readily satisfy this condition. Expectation values of other sparse observables can be estimated in the same way by replacing the Hamiltonian in Eq.~\eqref{eq:Eloc} by the observable of interest.

The optimal parameters approximating the ground state are those that minimize the energy expectation value $\langle H \rangle$.  We find them using Stochastic Reconfiguration (SR), a gradient-based method that accounts for the local geometry of the parameter manifold to make the optimization more efficient~\cite{carleo2017solving, hackl2020geometry, sorella2005wave}. More details can be found in the Appendix~\ref{app:optimization}.

We employ the Metropolis-Hastings algorithm~\cite{metropolis1953equation, hastings1970monte} to generate samples from the Born distribution defined on the composite Hilbert space $\mathcal H^{\text{photons}}\otimes\mathcal H^{\text{spins}}$. For a lattice of $L\times L$ sites, we use the following proposal distribution: with probability $1/L$, add or remove one photon (with equal probability), and with probability $(L-1)/L$, flip a randomly chosen spin. Further details and hyperparameters of the optimization are provided in Appendix~\ref{app:optimization}.

This approach allows us to obtain the ground state of the stoquastic Hamiltonian $\tilde H=U_aHU_a^\dagger$ with $U_a$ defined in Eq.~\eqref{eq:U_a}. To obtain the ground state of the original Hamiltonian $H$ of interest, we need to apply $U_a$ to the wavefunction. Provided a unitary $U$ is sparse in the basis $\{\ket{x}\}$, it is possible to evaluate $\bra{x}U\ket{\psi_{\boldsymbol\theta}}$ for any $\ket x$. This is sufficient to compute all quantities of interest through Monte Carlo sampling. This also allows us to apply symmetry operators to the NQS, and will be used for studying the properties of different symmetry sectors in Section~\ref{sec:physics}.

\subsection{Neural network architecture}
\label{sec:architecture}

We are now in a position to present the NQS ansatz that we introduce in this work. A central challenge in constructing an NQS for spin–photon systems is incorporating the infinite photon degree of freedom. Naively, one might pass the photon number $n$ as an additional input, together with the spin configuration. This approach, however, turned out inefficient in our exploratory experiments; our attempts with recurrent neural networks and convolutional neural networks led to numerical instabilities or insufficient variational energies. More details about these attempts can be found in the Appendix~\ref{app:nn}.

We instead introduce a multi-head architecture that treats the photon number $n$ in a fundamentally different way. As depicted in Fig.~\ref{fig:architecture}, the network $f_{\boldsymbol\theta}$ takes a spin configuration $\boldsymbol\sigma$ as input and produces, in its final layer, a vector with size $n_{\mathrm{cutoff}}+1$. Each component corresponds to the log-amplitude of the wavefunction for a given photon number $\log\psi_{\boldsymbol\theta}(n,\boldsymbol\sigma)$. In this way, the network simultaneously represents all photon sectors for a fixed spin configuration. In other words, the components of the log-wavefunction are $\log\psi_{\boldsymbol\theta}(n,\boldsymbol\sigma)=[f_{\boldsymbol\theta}(\boldsymbol\sigma)]_n$.  

This construction leaves the choice of the neural network architecture underlying the multi-head component free. In this work, we choose a ResNet, represented in Fig.~\ref{fig:architecture}, which has been established before as an efficient architecture for spin models~\cite{chen2024empowering}. It is made of convolutional operations that act locally on each element of the input and can be made translationally equivariant. This makes them well-suited for translationally symmetric two-dimensional spin Hamiltonians with local interactions. The residual connections skipping convolutional layers help mitigate the vanishing gradient problem encountered in deep neural networks.

The ResNet architecture used for spin-only systems outputs one wavefunction coefficient. It does this, after applying the final convolutional block, by averaging all the feature maps to one number, which, in the absence of the photon, represents the wavefunction amplitude of the input spin configuration. To accommodate the multi-head structure, we instead append to the final convolutional block an average pooling layer followed by a fully connected dense layer, represented as the last two layers in Fig.~\ref{fig:architecture}. The average pooling layer outputs a vector whose size is given by the number of feature maps in the convolutional layers. The dense layer then outputs a vector of size $n_{\mathrm{cutoff}}+1$, as required, enabling the network to simultaneously represent both spin and photon degrees of freedom. Further details about the architecture can be found in the Appendix~\ref{app:nn}.

\subsection{Symmetries of the NQS}

The multi-head architecture introduced in Section~\ref{sec:architecture} has the advantage that it can easily be made translationally invariant. We achieve this by ensuring that the convolution layers use a periodic padding, a stride of unity, and preserve the spatial dimensions of the input channels. This choice fixes the symmetry of the variational state; it represents, by construction, an eigenstate of the unit cell translation operator $\mathcal T$ with positive parity.

For the $\mathcal P$-symmetry on the other hand, the symmetry broken state with a negative order parameter $\qexp{a}/\sqrt N<0$ has negative wavefunction coefficients and therefore cannot be represented by the NQS. The other symmetry broken state with $\qexp{a}/\sqrt N>0$ and the symmetric state obeying $U_{\mathcal{P}}\ket{\psi_{\text{sym}}}=\ket{\psi_{\text{sym}}}$ are positive and can be represented by the NQS. Because $U_{\mathcal{P}}$ is diagonal in the computational basis with entries only $\pm 1$, it acts as $\bra{n,\boldsymbol{\sigma}}U_{\mathcal{P}}\ket{\psi}=\pm \braket{n,\boldsymbol{\sigma}}{\psi}$ for any configuration $\ket{n,\boldsymbol{\sigma}}$. The symmetric ground state $\ket{\psi_{\text{sym}}}$ must therefore have zero amplitude on configurations satisfying $\bra{n,\boldsymbol{\sigma}}U_{\mathcal{P}}\ket{n,\boldsymbol{\sigma}}=-1$. 

The initial state $\ket{\psi_0}$ does not have this structure, and the optimization therefore finds the symmetry-broken state, which also lacks this structure. The initial state, in our case, is obtained by sampling the weights of the neural network from a truncated normal distribution with small variances given in the Appendix~\ref{app:nn}. Physically, this initial network represents a state close to an equal superposition of all computational basis states, $\ket{\psi_0}\propto\sum_k\sum_{\mathbf s\in\{0,1\}^N}\ket{k}\otimes\ket{\sigma_1,\sigma_2,\cdots,\sigma_N}$.

\subsection{Enhancing the optimization}
\label{sec:optimization tricks}

\begin{figure}[]
\includegraphics[width=0.5\textwidth]{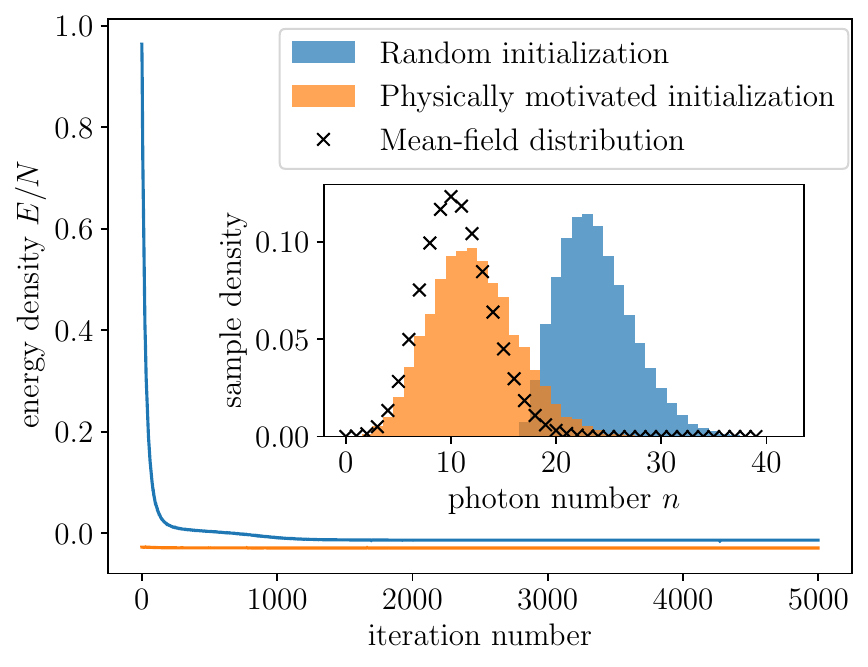}
\caption{\label{fig:trick} Comparison of the training curves between a state initialized with random weights and a state initialized as the ground state of the simplified Hamiltonian of Eq.~\eqref{eq:modified hamiltonian}. The physically motivated initialization reaches lower energies, illustrating the advantage of starting the optimization from a state closer to the ground state. The value chosen for the parameter $\alpha$ of the modified Hamiltonian in Eq.~\ref{eq:modified hamiltonian} is the average photon number of the mean-field ground state. The inset shows the photon distributions of each state at the end of the optimization. This illustrates how the improvement in energy can translate into significantly different values for the observables, such as the average photon number in this case. This example is for a $12\times12$ lattice with $g/\Delta=1.15$ and $V/\Delta=6.0$.
}
\end{figure}

In our simulations, we found that straightforward optimization starting from random network weights was sometimes unstable, could stall in local minima, and produced inaccurate observables. As explained in Section~\ref{sec:architecture}, small random weights correspond to a state close to an equal superposition of all computational basis states. This nonphysical state has an almost uniform photon distribution and lacks the correlations expected in the ground state. We observed that optimizations starting from such a state become increasingly problematic as the photon cutoff grows, particularly in regions where the photon number varies rapidly with the spin-photon coupling $g/\Delta$. This is illustrated in Fig.~\ref{fig:trick}, which compares training curves and photon distributions between the random initialization and a physically motivated initialization that we introduce in the rest of this section.

To construct a more suitable starting point, we introduce a modified Hamiltonian $\bar H$ that decouples spins and photons while retaining as much structure as possible of the original model. The energy of this new Hamiltonian is easier to minimize by design and its ground state can therefore serve as a better and more reliably obtainable starting point for the optimization. The modified Hamiltonian reads as
\begin{align}
\label{eq:modified hamiltonian}
    \bar H_{\text{}}(\alpha) = & -\frac{\Delta}{2} \sum_{\ell=1}^{N} \sigma_{\ell}^{z} + \frac{V}{d} \sum_{\langle \ell m \rangle} \left( \frac{1 + \sigma_{\ell}^{z}}{2} \right) \left( \frac{1 + \sigma_{m}^{z}}{2} \right)\nonumber\\
    & +\frac{2g\mathrm{Re}{(\alpha)}}{\sqrt{N}}\sum_{\ell=1}^{N} \sigma_{\ell}^{x}+\omega (a^\dagger-\alpha^*) (a-\alpha).
\end{align}
The first two terms act on the spins only and stay the same as in the original Hamiltonian. We replaced the creation and annihilation operators in the interaction term by the constant $\alpha\in\mathbb{C}$ to decouple spins and photons. This transformation is exact whenever the state of the system is a product state of the spin part and a coherent photon part $\ket{\psi}=\ket{\psi_{\mathrm{spin}}}\otimes\ket{\alpha}$. This state is close to the mean-field states used in Appendix~\ref{app:MF coherent}, the difference being that the spin part is not necessarily a product of individual spin states. This step helps to ensure that the ground state of $\bar H$ will at least be capable of capturing the mean-field physics, providing the physically motivated initialization we look for. The last term also contributes to this goal. It is a displaced Harmonic oscillator, and the only term acting on the photon degrees of freedom. Its energy is minimized by the coherent state $\ket{\alpha}$, allowing the photon part of the ground state to mimic the mean-field ground state.

We first optimize to find the ground state of $\bar H(\alpha)$. We then use the resulting NQS parameters to initialize the optimization for finding the ground state of the original Hamiltonian $H$. As illustrated in Fig.~\ref{fig:trick}, we find that this procedure leads to lower variational energies compared to a random initialization. The value of $\alpha$ can be chosen freely, with a natural choice given by the mean-field prediction. As shown in Fig.~\ref{fig:trick}, however, the photon distribution of the ground state deviates from mean-field, so choosing a different value of $\alpha$ can sometimes lead to lower variational energies. In practice, the values of $\alpha$ can be refined by scanning a small set of values and comparing the achieved variational energies.

Initializing an NQS with a mean-field result has been done for a Pfaffian-based architecture~\cite{chen2025neural}, where the mean-field state can be directly embedded without the need for optimization. For a general neural network ansatz, however, this is not possible, as there is no available mapping from a mean-field state to a specific set of weights. The approach we described instead provides an architecture-agnostic way to obtain a physically motivated initialization. Furthermore, finding an intermediate Hamiltonian like $\bar H$ can be done for any many-body Hamiltonian by following a similar procedure of decoupling subsystems and biasing them toward the mean-field solution.

In some cases, the optimization can also be improved via transfer learning~\cite{viteritti2026approaching, moss2025leveraging}: the neural network is first trained to find the ground state on a small lattice ($6\times6$ or $8\times8$) and then used to initialize the optimization for larger lattice sizes. This approach led to lower variational energies in regions with low spin–photon coupling $g/\Delta$, where the ground state contains very few photons.

\begin{figure}
\includegraphics[width=0.5\textwidth]{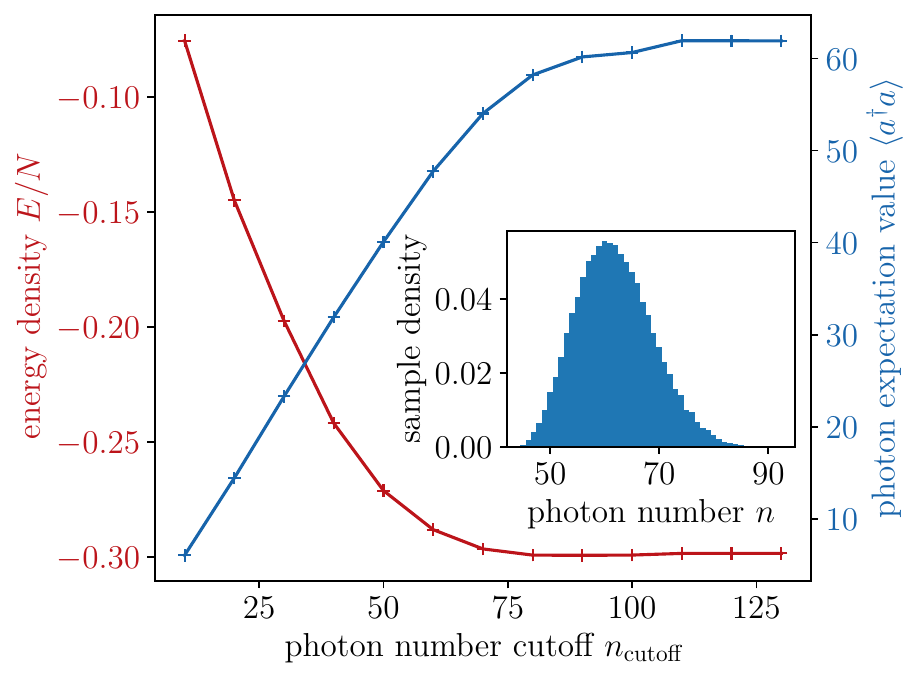}
\caption{\label{fig:cutoff} Scaling of the average photon number $\qexp{a^\dagger a}$ and the energy density as the photon cutoff is increased. The cutoff must be chosen so that, if it is increased, the photon number in the cavity does not change. The inset shows the photon distribution of the ground state at this point. This example is for a $12\times12$ lattice at $g/\Delta=1.5$ and $V/\Delta=6$. The error bars on the main graph are too small to be visible.
}
\end{figure}

\begin{figure*}[t]
    \includegraphics[width=\textwidth]{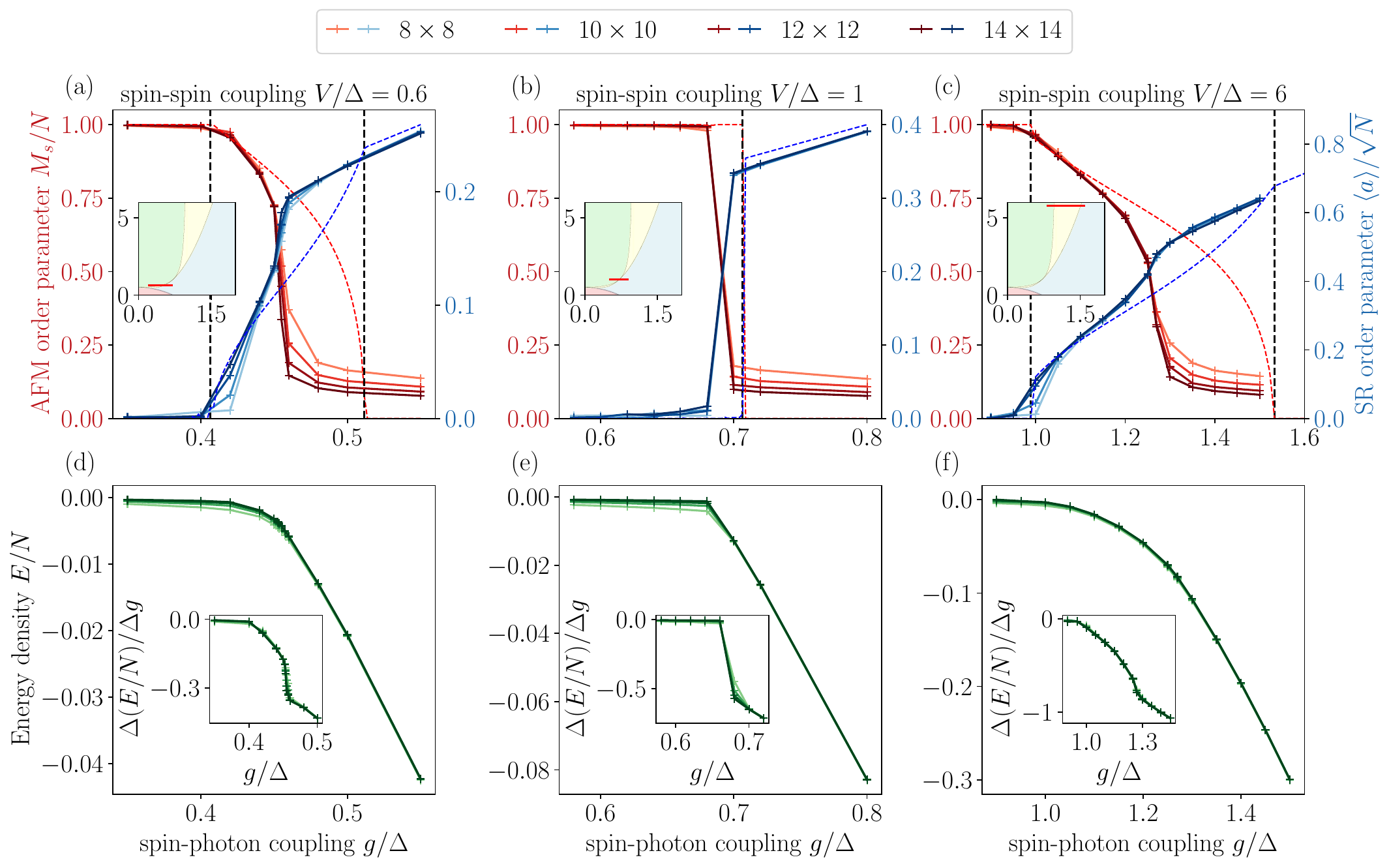}
    \caption{\label{fig:cuts} (a-c) AFM and SR order parameters showcasing the different phase transitions for various cuts through the phase diagram at $V=0.6$, $V=1$, and $V=6$. The dashed red and blue curves represent the mean-field predictions for the two quantities. The black dashed vertical lines show the position of the phase transitions according to mean-field. (d-f) The second row shows the corresponding energy densities and their derivative with respect to $g/\Delta$. In the thermodynamic limit, a first order phase transition is characterized by a kink in the energy density and a discontinuity in its derivative. The error bars are too small to be visible. Details about the cutoffs chosen for each data point can be found in Appendix~\ref{app:optimization}.
    }
\end{figure*}

\subsection{Photon number cutoff}
\label{sec:cutoff}

The photon cutoff $n_{\mathrm{cutoff}}$ determines the dimension of the network output and limits the maximal photon number represented. Since the ground-state photon distribution is not known a priori, the cutoff must be validated a posteriori. For this purpose, we increase $n_{\mathrm{cutoff}}$ and re-optimize; if the average photon number and variational energy remain unchanged, the cutoff is sufficient. An example is shown in Fig.~\ref{fig:cutoff}, where the photon number saturates beyond a certain cutoff. Note that the cutoff is substantially larger than the average photon number, because in this case, the photon distribution is spread out as shown in the inset of Fig.~\ref{fig:cutoff}. Furthermore, by virtue of the variational principle, the ground state energy in Fig.~\ref{fig:cutoff} decreases as the cutoff grows before stabilizing when a sufficiently high cutoff is reached.

We emphasize that the initialization strategy described in Section~\ref{sec:optimization tricks} becomes increasingly important at large cutoff values, where uniform initial photon distributions lead to numerical instabilities. In some regimes, the photon distribution is concentrated around a nonzero value, as in Fig.~\ref{fig:cutoff}. In such cases, introducing a lower photon bound can further improve convergence. As for the upper cutoff, we chose the lower cutoff by decreasing the bound until observables cease to change. These modifications become crucial for large system sizes that contain more photons or for states deeper in the superradiant phase.

\section{Results and Discussion}
\label{sec:results}

\subsection{Phase boundaries}

With the method now established, we can start exploring the physical properties predicted by NQS.
To probe the phase diagram, we consider three representative cuts. In the first region, we fix the spin-spin coupling to $V/\Delta=0.6$, while varying the spin-photon coupling $g/\Delta$ to explore all phase transitions predicted by mean-field theory. The AFM order parameter and the photon density along this cut are shown in Fig.~\ref{fig:cuts}(a), together with the mean-field predictions shown as dashed lines. The lattice sizes range from $8\times 8$ to $14\times 14$, and the maximal photon number cutoff used is 120. This cutoff is needed to represent the superradiant ground states for the largest system sizes, which can host more than a hundred photons.

For small values of $g/\Delta$, the NQS results closely follow the mean-field predictions.
The AFM-to-mixed phase transition occurs for $g/\Delta\in[0.4,0.42]$, in agreement with the mean-field result indicated by the left vertical dashed line. This observation is consistent with previous work demonstrating that the mean-field description of the AFM-to-mixed phase transition is exact in the thermodynamic limit \cite{schellenberger2024almost, schneider2024dipolar}. At larger values of $g/\Delta$, we observe a deviation: the mixed-to-SR phase transition occurs for $g/\Delta\in[0.45,0.46]$, earlier than predicted by mean-field theory. The mixed phase is therefore present, but over a smaller range of spin-photon coupling.

The second region we explore is another cut for a fixed spin-spin coupling of $V/\Delta=1$. It is of particular interest because mean-field theory predicts the presence of a multi-critical point at which the mixed phase vanishes. Our results, shown in Fig.~\ref{fig:cuts}b, are consistent with this prediction: within the limits of our resolution along the $g/\Delta$ axis, we do not observe a mixed phase. Both the AFM and SR order parameters display transitions that appear to be first order, in agreement with mean-field theory. The transition is, however, shifted to smaller values of $g/\Delta$, consistent with previous numerical results~\cite{langheld2025quantum}. We note that the results of Refs~\cite{schellenberger2024almost, schneider2024dipolar} are not applicable in this case, as they rely on the low photon occupation limit, which is not satisfied for a first order phase transition.

Finally, we consider a third cut at $V/\Delta=6$. In this regime, mean-field theory predicts a significantly larger mixed phase than at smaller spin-spin couplings. 
This region's interest lies in the challenge that it presents for our method; it requires exploring larger values of the spin-photon coupling $g/\Delta$, which leads to a larger number of photons in the ground state of the SR phase. The results, shown in Fig.~\ref{fig:cuts}(c), lead to the same conclusion as for the $V/\Delta=0.6$ region: the mixed phase is again smaller than predicted by mean-field theory.

Figures~\ref{fig:cuts}(d-f) show the ground state energy along the three cuts considered. These curves can give indications about the nature of the phase transitions. In the thermodynamic limit, the presence of a kink would signal a level crossing and therefore a first-order phase transition. Equivalently, such a transition would appear as a discontinuity in the derivative of the energy with respect to $g/\Delta$. Finite system sizes, however, prevent the direct observation of a kink and make the interpretation difficult. We nonetheless clearly observe a behavior consistent with a first-order transition at $V/\Delta=1$.

\subsection{Spin-photon correlations}
\label{sec:physics}

\begin{figure}[]
\includegraphics[width=0.5\textwidth]{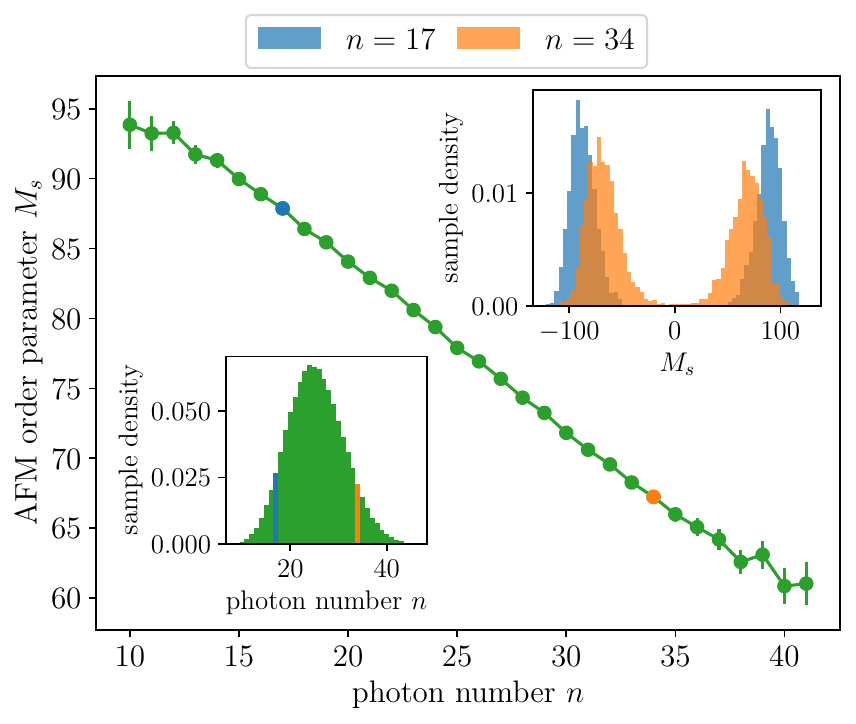}
\caption{\label{fig:entanglement} Illustration of the correlations between spins and photons captured by the NQS approach. The NQS state is sampled, giving the photon distribution in the bottom left inset. For each sample with a given photon number $n$, the AFM order parameter of the spin counterpart is computed and averaged according to Eq.~\eqref{eq:Msn}. Repeating this procedure for each photon number gives the main curve, which shows a correlation between photon occupation and antiferromagnetic order. It indicates that higher photon occupation suppresses the AFM spin ordering. The top-right inset shows the distribution of AFM order parameters for two representative photon numbers, illustrating the reduced AFM order of configurations with larger photon numbers. The system used has a $12\times12$ lattice with $V/\Delta=6$ and $g/\Delta=1.25$.
}
\end{figure}

The deviations from mean-field theory observed in Fig.~\ref{fig:cuts} can be attributed to the ability of NQS to capture the correlations present in the ground state. These correlations play an important role in the mixed phase, where the discrepancies are the largest. To gain further insight, we now investigate the correlations between the spins and the photons.

We first introduce a new quantity that resolves the AFM order parameter as a function of the photon number. Given a set of configurations $\{\ket{n_k, \boldsymbol \sigma_k}\}$ sampled from the Born distribution, with $\boldsymbol \sigma_k=(\sigma_{k,1}, \sigma_{k,2},\dots,\sigma_{k,N})$ the spin configuration, we define 
\begin{equation}
\label{eq:Msn}
    M_s(n)=\frac{1}{\sum_k\delta_{n_k n}}\sum_kM_s(n, \boldsymbol \sigma_k)\delta_{n_k n},
\end{equation}
where $M_s(n, \boldsymbol  \sigma_k)=\qexp{\boldsymbol  \sigma_{k,\text{even}}^z}-\qexp{\boldsymbol  \sigma_{k,\text{odd}}^z}$ is the staggered magnetization for a single configuration. We plot this quantity as a function of the photon number $n$ in Fig.~\ref{fig:entanglement} for a state in the mixed phase. 

This reveals that, within a given state, the AFM order is stronger for low photon numbers and gradually decreases as the photon number increases. Physically, this can be understood by noting that the spin-photon coupling acts similarly to a transverse field. The $\sigma^x$ term tends to align the spins along the $x$-axis, suppressing the AFM order, and ultimately leads to the SR phase. As the photon number in the cavity increases, this effect becomes stronger.

The curve in Fig.~\ref{fig:entanglement} is approximately linear with a slope close to $-1$, suggesting that adding one photon reduces the AFM order parameter by one. Given a configuration $\{\ket{n_k, \boldsymbol \sigma_k}\}$, flipping a spin in $\boldsymbol \sigma_k$ changes the AFM order parameter $M_s(n, \boldsymbol  \sigma_k)$ by $\pm 2$. To get a change of $\pm 1$ in the AFM order parameter, the spin would instead need to become an equal superposition of the up and down states. This suggests the following interpretation: each additional photon effectively forces one spin previously aligned along the $z$ axis to align along the $x$ axis, leading to the observed reduction in AFM order. These spin-photon correlations cannot be captured by mean-field theory and explain the deviation with respect to the NQS results shown in Fig.~\ref{fig:cuts}(c). Having explored the spin-photon correlations, we now turn to the entanglement between those two subsystems.

\begin{figure*}[t]
\includegraphics[width=1\textwidth]{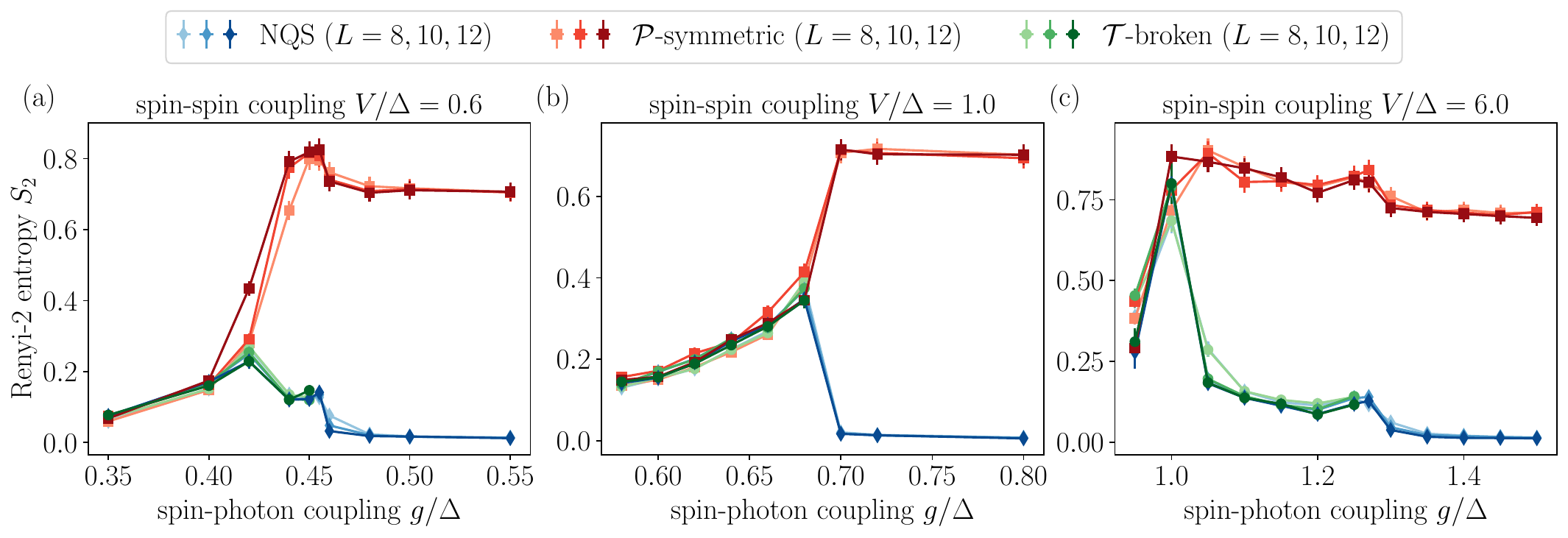}
\caption{\label{fig:renyi} Renyi-2 entanglement entropy for the same three cuts as in Fig.~\ref{fig:cuts}, at (a) $V=0.6$, (b) $V=1$, and (c) $V=6$, for lattice sizes $L=8,10,12$. In the AFM phase, the SR symmetry is not broken, so symmetrizing the ground state has no effect, the NQS (blue) and $\mathcal P$-symmetric (red) curves overlap. After the AFM-to-mixed phase transition, however, the NQS finds a ground state with the SR symmetry broken. Symmetrizing it (red curves) leads to a larger entanglement entropy. On the other hand, the NQS result is always $\mathcal T$ symmetric, but breaking this symmetry does not change the Renyi-2 entanglement entropy, as shown by the $\mathcal T$ broken (green) and NQS (blue) lines overlapping. After the mixed-to-SR phase transition, the translation symmetry can no longer be broken, so the green curve stops.
}
\end{figure*}

\subsection{Spin-photon entanglement}

To go further beyond the mean field solution, and to take advantage of the NQS versatility, we compute the Renyi-2 entanglement entropy $S_2$ between the spin lattice and the cavity. It quantifies the entanglement between the spin and photon subsystems and is therefore a useful way to study effects of strong correlations. It is defined as
\begin{equation}
    S_2=-\ln(\tr \rho_{\mathrm{photon}}^2)
\end{equation}
where $\rho_{\mathrm{photon}}=\tr_{\mathrm{spin}} \left[\ket{\psi}\bra{\psi}\right]$ is the reduced density matrix of the photon subsystem obtained by tracing out the spin subsystem. The purity of $\rho_{\mathrm{photon}}$ can be estimated via Monte-Carlo sampling as~\cite{hastings2010measuring, shi2023measuring}
\begin{equation}
\label{eq:purity}
    \tr\rho_{\mathrm{photon}}^2=\expval{\frac{\psi(n',\boldsymbol \sigma)\psi(n,\boldsymbol \sigma')}{\psi(n,\boldsymbol \sigma)\psi(n',\boldsymbol \sigma')}}{\pi,\pi'}.
\end{equation}
The samples $\{n,\boldsymbol \sigma\}$ and $\{n',\boldsymbol \sigma'\}$ are drawn independently from the Born distributions $\pi,\pi'$ defined in Section~\ref{sec:sr}. The configurations in the denominator are obtained by swapping the photon occupation numbers of the sampled pairs of configurations, giving $\{n',\boldsymbol \sigma\}$ and $\{n,\boldsymbol \sigma'\}$. If there is no entanglement, the wavefunction factorizes, and the quantity inside of the expectation value in Eq.~\eqref{eq:purity} will consistently give unity. Therefore, intuitively, the swapping quantifies how close $\psi$ is to a product state, and any deviation from unity will signal the presence of entanglement.

We show the Renyi-2 entanglement entropy in Fig.~\ref{fig:renyi} for the same cuts through the phase diagram as in Fig.~\ref{fig:cuts}. Notice that the Renyi-2 entanglement entropy can take different values across states within the ground state manifold. We show its value in different symmetry sectors by either symmetrizing the NQS ground state with respect to the $\mathcal P$-symmetry, or by breaking its $\mathcal T$-symmetry.

As discussed in Section~\ref{sec:architecture}, the ground state found by the NQS is always translationally symmetric. In the regions where the unit cell translation symmetry $\mathcal T$ is spontaneously broken, the NQS ground state is therefore a positive eigenstate of $U_\mathcal T$. Given the symmetric wave function $\psi_S(n,\boldsymbol\sigma)$, a symmetry-broken wave function is obtained via $\psi_{SB}(n,\boldsymbol\sigma)=\psi_S(n,\boldsymbol\sigma)\theta({M_s(n,\boldsymbol\sigma))}$ with $M_s(n,\boldsymbol\sigma)$ defined in Eq.~\eqref{eq:Msn}, and $\theta$ the Heaviside step function selecting positive values of $M_s(n,\boldsymbol\sigma)$. This uses the fact that the AFM order parameters of single configurations $\ket{n,\boldsymbol\sigma}$ have a different sign in each symmetry sector. The Renyi-2 entanglement entropy of the resulting state with its symmetry broken is shown using green colors in Fig.~\ref{fig:renyi}. We observe that its entanglement entropy does not change with respect to that of the original unit cell translation symmetric state $\psi(n, \boldsymbol{\sigma})$ (in blue). This indicates that the photon mode is independent of the sign of the AFM order parameter. Instead, it depends only on its magnitude, consistent with the behavior observed in Fig.~\ref{fig:entanglement}.

On the other hand, when the ground state is superradiant, the NQS finds a non-symmetric state with positive order parameter $\qexp{a}/\sqrt N>0$; the corresponding symmetrized state is $\ket{\psi_{\text{sym}}}=(\ket{\psi}+U_{\mathcal{P}}\ket{\psi})/\sqrt 2$, with $U_{\mathcal{P}}$ defined in Eq.~\eqref{eq:U_sr}. The Renyi-2 entanglement entropy of this state is shown in Fig.~\ref{fig:renyi} as the orange curve. In the region where the symmetry is spontaneously broken, the entanglement entropy increases roughly by a constant offset of $\ln(2)$, indicating that the spin state depends on the sign of the SR order parameter. To sample the $\mathcal P$-symmetric state and compute its entropy, we had to modify the proposal distribution described in Section~\ref{sec:sr}. Because the $\mathcal P$ symmetry conserves the parity of the total number of excitations, we had to add the possibility to flip a spin and add or remove a photon at the same time.

The Renyi-2 entanglement entropy in Fig.~\ref{fig:renyi} shows a sudden change of behavior at the AFM-to-mixed phase transition. To the left of this transition, for weaker spin-photon couplings $g/\Delta$, the entanglement entropy gradually increases, indicating that the antiferromagnet is increasingly dressed by spin-photon fluctuations. At the transition, it then sharply decreases or increases depending on whether the ground state is $\mathcal P$-symmetric or not. For the cuts at $V/\Delta=0.6$ and $V/\Delta=6$, the mixed-to-SR phase transition is marked by a small peak in the entanglement entropy. Then, at stronger spin-photon coupling, in the superradiant phase, the entanglement vanishes. To understand this behavior, it is useful to complete the squares in the Hamiltonian given in Eq.~\eqref{eq:full H}, yielding:
\begin{align}
    H=&H_{\mathrm{spin}}-\frac{g^2}{\omega_0 N}\left(\sum_{l=1}^N\sigma_l^x\right)^2\\
    &+\omega_0\left(a^\dagger+\frac{g}{\omega_0\sqrt N}\sum_{l=1}^N\sigma_l^x\right)\left(a+\frac{g}{\omega_0\sqrt N}\sum_{l=1}^N\sigma_l^x\right).\nonumber
\end{align}
For $g/\Delta\gg1$, the energy of $H_{\mathrm{spin}}$ is negligible. The energy of the second term, an all-to-all $\sigma^x$ coupling, is minimized when the spins are all aligned along $x$. As for the spin-photon coupling term, if the spins are in an eigenstate of $\sigma^x$, it takes the form of a displaced oscillator whose energy is minimized to $0$ when the photons are in a coherent state. The energy of this last quadratic term cannot be negative, meaning that no change to the photons and no spin-photon entanglement can further decrease the energy of this term. For large spin-photon coupling $g/\Delta$, the ground state therefore tends toward a product state, explaining the lack of entanglement.

Overall, $\mathcal{P}$-symmetry breaking has the strongest effect on the entanglement properties. We do not observe, however, any system size dependent scaling of the entropy. Indeed, due to the homogeneous coupling of the photon mode to the spin degrees of freedom, correlations can only relate to permutation invariant properties of the spin system, limiting the total entanglement to sub-extensive values.

\section{Conclusion \& Outlook}

In this work, we have introduced a multi-head neural network architecture capable of describing strongly interacting spin-boson systems. It efficiently represents wavefunctions on hybrid Hilbert spaces involving infinite-dimensional bosonic degrees of freedom by facilitating very large occupation number truncations for accurate approximation. Thereby, it can handle large system sizes beyond the reach of exact diagonalization.

We have applied this framework to a two-dimensional array of Rydberg atoms coupled to a cavity mode. The neural quantum state approach has allowed us to take correlations into account and probe the ground state properties beyond mean-field. Our results confirm the qualitative structure of the mean-field phase diagram, including the existence of a mixed phase in which antiferromagnetic order coexists with superradiance. Quantitatively, however, we find shifted phase boundaries, resulting in a mixed phase that is smaller than predicted by mean-field theory.

Beyond locating phase boundaries, the method provides direct access to ground-state correlations. In the mixed phase, we observe clear spins-photons correlations that contribute to the deviation from mean-field predictions. We have further estimated the Renyi-2 entanglement entropy between spins and photons across the ground state manifold and have found that the superradiant related $\mathcal P$-symmetry provides the dominant contribution to the entanglement. We furthermore observe an intensive scaling of the entanglement entropy with system size, consistent with the restriction imposed by the homogeneous spin-photon coupling.

The multi-head neural-network architecture is a key ingredient of this work. This particular treatment of the bosonic degree of freedom is motivated by convergence and scaling challenges that arise when straightforwardly including the photon number as a network input. The optimization procedure also plays a crucial role: introducing a physically motivated intermediate Hamiltonian provides an initialization that significantly improves convergence. This strategy is general and can be applied to a wide range of variational problems.

In addition to the multi-head NQS, alternative architectures may offer similar or even superior performance for problems defined on hybrid Hilbert spaces. For instance, a recent work has demonstrated the viability of a transformer-based ans\"atz for spin-fermion systems~\cite{rende2026transformer}. Exploring a broader class of architectures may therefore  further improve the accuracy, efficiency and scalability of such methods.

The NQS approach presented here is well-suited for two-dimensional systems, where tensor network methods face limitations~\cite{white1992density, deng2017quantum}. That said, the model considered in this work is sign-free and therefore within reach of quantum Monte Carlo (QMC) methods. Neural quantum states, however, do not experience the sign problem that hinders QMC and have enabled the study of non-stoquastic systems~\cite{troyer2005computational, chen2024empowering, bukov2021learning}. For instance, state-of-the-art variational energies have been achieved for the frustrated $J_1$-$J_2$ Heisenberg model on the square lattice, where frustrations give rise to a non-trivial ground state sign structure~\cite{chen2024empowering}. We therefore anticipate that the NQS approach will open the way to the study of more complex systems. In particular, recent proposals suggest that Rydberg arrays in cavities could be used to study spin liquids~\cite{mann2025squeezing}, providing a concrete setting where NQS methods may offer a unique advantage.

More broadly, the approach is not limited to the model studied here. Its ability to handle hybrid Hilbert spaces where one constituent degree of freedom has an infinite local dimension makes it applicable to a large variety of systems. This includes spin–phonon models and lattice gauge theories with truncated gauge fields~\cite{mahajan2025structure, orbach1961spin,Chandrasekharan1997}. The approach is also amenable to a straightforward generalization to Floquet systems formulated in the extended Hilbert space formalism~\cite{sambe1973steady}. Indeed, in this formulation, physical states gain an additional infinite-dimensional degree of freedom similar to that of a bosonic mode. This could therefore provide a way to study non-equilibrium quantum dynamics with the same multi-head architecture.

Overall, this work demonstrates the potential of neural quantum states to become a powerful tool for investigating strongly correlated quantum systems with both spin and bosonic degrees of freedom, beyond the capabilities of existing numerical methods.

\begin{acknowledgements}
We acknowledge fruitful discussions with Achim Rosch, Jamir Marino, Johannes Hofmann, Jonas Rigo, and Wladislaw Krinitsin. 
Funded by the Deutsche Forschungsgemeinschaft (DFG, German Research
Foundation) -- 544919793, and the European Union (ERC, QuSimCtrl, 101113633). Views and opinions expressed are however those of the authors only and do not necessarily reflect those of the European Union or the European Research Council Executive Agency. Neither the European Union nor the granting authority can be held responsible for them.
MS was supported through the Helmholtz Initiative and Networking Fund, Grant No.~VH-NG-1711.
The numerical simulations were implemented using the jVMC codebase \cite{jvmc2022}.
The authors gratefully acknowledge the scientific support and HPC resources provided by the Erlangen National High Performance Computing Center (NHR@FAU) of the Friedrich-Alexander-Universität Erlangen-Nürnberg (FAU) under the NHR project b242da / JA-22808. NHR funding is provided by federal and Bavarian state authorities.
Numerical simulations were also performed on the MPCDF HPC cluster.

\textit{Data and code availability.}---The data, code, and trained models are available under DOI: \href{https://doi.org/10.5281/zenodo.20630702}{10.5281/zenodo.20630702.}

\end{acknowledgements}

\bibliography{bibliography}

@article{gelhausen2016quantum,
  title={Quantum-optical magnets with competing short-and long-range interactions: Rydberg-dressed spin lattice in an optical cavity},
  author={Gelhausen, Jan and Buchhold, Michael and Rosch, Achim and Strack, Philipp},
  journal={SciPost Physics},
  volume={1},
  number={1},
  pages={004},
  year={2016},
  url={https://www.scipost.org/10.21468/SciPostPhys.1.1.004?acad_field_slug=politicalscience}
}

@article{bezvershenko2021dicke,
  title = {Dicke Transition in Open Many-Body Systems Determined by Fluctuation Effects},
  author = {Bezvershenko, Alla V. and Halati, Catalin-Mihai and Sheikhan, Ameneh and Kollath, Corinna and Rosch, Achim},
  journal = {Phys. Rev. Lett.},
  volume = {127},
  issue = {17},
  pages = {173606},
  numpages = {6},
  year = {2021},
  month = {Oct},
  publisher = {American Physical Society},
  doi = {10.1103/PhysRevLett.127.173606},
  url = {https://link.aps.org/doi/10.1103/PhysRevLett.127.173606}
}

@article{schauss2012observation,
  title={Observation of spatially ordered structures in a two-dimensional Rydberg gas},
  author={Schau{\ss}, Peter and Cheneau, Marc and Endres, Manuel and Fukuhara, Takeshi and Hild, Sebastian and Omran, Ahmed and Pohl, Thomas and Gross, Christian and Kuhr, Stefan and Bloch, Immanuel},
  journal={Nature},
  volume={491},
  number={7422},
  pages={87--91},
  year={2012},
  publisher={Nature Publishing Group UK London},
  url={https://www.nature.com/articles/nature11596}
}

@article{troyer2005computational,
  title={Computational Complexity and Fundamental Limitations to Fermionic Quantum Monte Carlo Simulations},
  author={Troyer, Matthias and Wiese, Uwe-Jens},
  journal={Physical Review Letters},
  volume={94},
  number={17},
  pages={170201},
  year={2005},
  publisher={APS},
  url={https://journals.aps.org/prl/abstract/10.1103/PhysRevLett.94.170201}
}

@article{bravyi2014monte,
  title={Monte Carlo simulation of stoquastic Hamiltonians},
  author={Bravyi, Sergey},
  journal={arXiv:1402.2295},
  year={2014},
  url={https://arxiv.org/abs/1402.2295}
}

@article{dicke1954coherence,
  title={Coherence in spontaneous radiation processes},
  author={Dicke, Robert H},
  journal={Physical Review},
  volume={93},
  number={1},
  pages={99},
  year={1954},
  publisher={APS},
  url={https://journals.aps.org/pr/abstract/10.1103/PhysRev.93.99}
}

@article{shi2023measuring,
  title={Measuring Renyi Entropy in Neural Network Quantum States},
  author={Shi, Han-Qing and Zhang, Hai-Qing},
  journal={arXiv:2308.05513},
  year={2023},
  url={https://arxiv.org/abs/2308.05513}
}

@article{carleo2017solving,
  title={Solving the quantum many-body problem with artificial neural networks},
  author={Carleo, Giuseppe and Troyer, Matthias},
  journal={Science},
  volume={355},
  number={6325},
  pages={602--606},
  year={2017},
  publisher={American Association for the Advancement of Science},
  url={https://www.science.org/doi/full/10.1126/science.aag2302}
}

@article{hackl2020geometry,
  title={Geometry of variational methods: dynamics of closed quantum systems},
  author={Hackl, Lucas and Guaita, Tommaso and Shi, Tao and Haegeman, Jutho and Demler, Eugene and Cirac, J Ignacio},
  journal={SciPost Physics},
  volume={9},
  number={4},
  pages={048},
  year={2020},
  url={https://scipost.org/SciPostPhys.9.4.048}
}

@article{chen2024empowering,
  title={Empowering deep neural quantum states through efficient optimization},
  author={Chen, Ao and Heyl, Markus},
  journal={Nature Physics},
  volume={20},
  number={9},
  pages={1476--1481},
  year={2024},
  publisher={Nature Publishing Group UK London},
  url={https://www.nature.com/articles/s41567-024-02566-1}
}

@article{de2026realization,
  title={Realization of a cavity-coupled Rydberg array},
  author={De Santis, Jacopo and Dura-Kov{\'a}cs, Bal{\'a}zs and {\"O}nc{\"u}, Mehmet and Bouscal, Adrien and Vasileiadis, Dimitrios and Zeiher, Johannes},
  journal={arXiv:2602.12152},
  year={2026},
  url={https://arxiv.org/abs/2602.12152}
}

@article{sorella2005wave,
  title={Wave function optimization in the variational Monte Carlo method},
  author={Sorella, Sandro},
  journal={Physical Review B},
  volume={71},
  number={24},
  pages={241103},
  year={2005},
  publisher={APS},
  url={https://journals.aps.org/prb/abstract/10.1103/PhysRevB.71.241103}
}

@article{hastings2010measuring,
  title={Measuring Renyi entanglement entropy in quantum Monte Carlo simulations},
  author={Hastings, Matthew B and Gonz{\'a}lez, Iv{\'a}n and Kallin, Ann B and Melko, Roger G},
  journal={Physical Review Letters},
  volume={104},
  number={15},
  pages={157201},
  year={2010},
  publisher={APS},
  url={https://journals.aps.org/prl/abstract/10.1103/PhysRevLett.104.157201}
}

@article{white1992density,
  title={Density matrix formulation for quantum renormalization groups},
  author={White, Steven R},
  journal={Physical Review Letters},
  volume={69},
  number={19},
  pages={2863},
  year={1992},
  publisher={APS},
  url={https://journals.aps.org/prl/abstract/10.1103/PhysRevLett.69.2863}
}

@article{deng2017quantum,
  title={Quantum entanglement in neural network states},
  author={Deng, Dong-Ling and Li, Xiaopeng and Das Sarma, S},
  journal={Physical Review X},
  volume={7},
  number={2},
  pages={021021},
  year={2017},
  publisher={APS},
  url={https://journals.aps.org/prx/abstract/10.1103/PhysRevX.7.021021}
}

@article{mahajan2025structure,
  title={Structure and dynamics of electron-phonon coupled systems using neural quantum states},
  author={Mahajan, Ankit and Robinson, Paul J and Lee, Joonho and Reichman, David R},
  journal={Physical Review B},
  volume={112},
  number={18},
  pages={184310},
  year={2025},
  publisher={APS},
  url={https://journals.aps.org/prb/abstract/10.1103/p8qz-p45d}
}

@article{orbach1961spin,
  title={Spin-lattice relaxation in rare-earth salts},
  author={Orbach, R},
  journal={Proceedings of the Royal Society of London. Series A. Mathematical and Physical Sciences},
  volume={264},
  number={1319},
  pages={458--484},
  year={1961},
  publisher={The Royal Society London},
  url={https://royalsocietypublishing.org/rspa/article-abstract/264/1319/458/11095/Spin-lattice-relaxation-in-rare-earth-salts?redirectedFrom=fulltext}
}

@article{sambe1973steady,
  title={Steady states and quasienergies of a quantum-mechanical system in an oscillating field},
  author={Sambe, Hideo},
  journal={Physical Review A},
  volume={7},
  number={6},
  pages={2203},
  year={1973},
  publisher={APS},
  url={https://journals.aps.org/pra/abstract/10.1103/PhysRevA.7.2203}
}

@article{browaeys2020many,
  title={Many-body physics with individually controlled Rydberg atoms},
  author={Browaeys, Antoine and Lahaye, Thierry},
  journal={Nature Physics},
  volume={16},
  number={2},
  pages={132--142},
  year={2020},
  publisher={Nature Publishing Group UK London},
  url={https://www.nature.com/articles/s41567-019-0733-z}
}

@article{bloch2012quantum,
  title={Quantum simulations with ultracold quantum gases},
  author={Bloch, Immanuel and Dalibard, Jean and Nascimbene, Sylvain},
  journal={Nature Physics},
  volume={8},
  number={4},
  pages={267--276},
  year={2012},
  publisher={Nature Publishing Group},
  url={https://www.nature.com/articles/nphys2259}
}

@article{landig2016quantum,
  title={Quantum phases from competing short-and long-range interactions in an optical lattice},
  author={Landig, Renate and Hruby, Lorenz and Dogra, Nishant and Landini, Manuele and Mottl, Rafael and Donner, Tobias and Esslinger, Tilman},
  journal={Nature},
  volume={532},
  number={7600},
  pages={476--479},
  year={2016},
  publisher={Nature Publishing Group UK London},
  url={https://www.nature.com/articles/nature17409}
}

@article{klinder2015observation,
  title={Observation of a superradiant Mott insulator in the Dicke-Hubbard model},
  author={Klinder, Jens and Ke{\ss}ler, Hans and Bakhtiari, M Reza and Thorwart, M and Hemmerich, Andreas},
  journal={Physical review letters},
  volume={115},
  number={23},
  pages={230403},
  year={2015},
  publisher={APS},
  url={https://journals.aps.org/prl/abstract/10.1103/PhysRevLett.115.230403}
}

@article{defenu2023long,
  title={Long-range interacting quantum systems},
  author={Defenu, Nicolo and Donner, Tobias and Macr{\`\i}, Tommaso and Pagano, Guido and Ruffo, Stefano and Trombettoni, Andrea},
  journal={Reviews of Modern Physics},
  volume={95},
  number={3},
  pages={035002},
  year={2023},
  publisher={APS},
  url={https://journals.aps.org/rmp/abstract/10.1103/RevModPhys.95.035002}
}

@inproceedings{he2015delving,
  title={Delving deep into rectifiers: Surpassing human-level performance on imagenet classification},
  author={He, Kaiming and Zhang, Xiangyu and Ren, Shaoqing and Sun, Jian},
  booktitle={Proceedings of the IEEE international conference on computer vision},
  pages={1026--1034},
  year={2015},
  url={https://openaccess.thecvf.com/content_iccv_2015/html/He_Delving_Deep_into_ICCV_2015_paper.html}
}

@article{hendrycks2016gaussian,
  title={Gaussian error linear units (gelus)},
  author={Hendrycks, Dan and Gimpel, Kevin},
  journal={arXiv:1606.08415},
  year={2016},
  url={https://arxiv.org/abs/1606.08415}
}

@article{rende2024simple,
  title={A simple linear algebra identity to optimize large-scale neural network quantum states},
  author={Rende, Riccardo and Viteritti, Luciano Loris and Bardone, Lorenzo and Becca, Federico and Goldt, Sebastian},
  journal={Communications Physics},
  volume={7},
  number={1},
  pages={260},
  year={2024},
  publisher={Nature Publishing Group UK London},
  url={https://www.nature.com/articles/s42005-024-01732-4}
}

@article{schellenberger2024almost,
  title={(Almost) everything is a Dicke model-Mapping non-superradiant correlated light-matter systems to the exactly solvable Dicke model},
  author={Schellenberger, Andreas and Schmidt, Kai Phillip},
  journal={SciPost Physics Core},
  volume={7},
  number={3},
  pages={038},
  year={2024},
  url={https://scipost.org/SciPostPhysCore.7.3.038}
}

@article{bukov2021learning,
  title={Learning the ground state of a non-stoquastic quantum Hamiltonian in a rugged neural network landscape},
  author={Bukov, Marin and Schmitt, Markus and Dupont, Maxime},
  journal={SciPost Physics},
  volume={10},
  number={6},
  pages={147},
  year={2021},
  url={https://scipost.org/SciPostPhys.10.6.147}
}

@article{marsh2025multimode,
  title={Multimode Cavity QED Ising Spin Glass},
  author={Marsh, Brendan P and Schuller, David Atri and Ji, Yunpeng and Hunt, Henry S and Socolof, Giulia Z and Bowman, Deven P and Keeling, Jonathan and Lev, Benjamin L},
  journal={Physical Review Letters},
  volume={135},
  number={16},
  pages={160403},
  year={2025},
  publisher={APS},
  url={https://journals.aps.org/prl/abstract/10.1103/x19r-pzyb}
}

@article{an2022quantum,
  title={Quantum phase transition of the two-dimensional Rydberg atom array in an optical cavity},
  author={An, Gao-Qi and Wang, Tao and Zhang, Xue-Feng},
  journal={arXiv:2204.08800},
  url={https://arxiv.org/abs/2204.08800},
  year={2022}
}

@article{langheld2025quantum,
  title={Quantum phase diagrams of Dicke-Ising models by a wormhole algorithm},
  author={Langheld, Anja and H{\"o}rmann, Max and Schmidt, Kai Phillip},
  journal={Physical Review B},
  volume={112},
  number={16},
  pages={L161123},
  year={2025},
  publisher={APS},
  url={https://journals.aps.org/prb/abstract/10.1103/PhysRevB.106.134506}
}

@article{assaraf1999zero,
  title={Zero-variance principle for Monte Carlo algorithms},
  author={Assaraf, Roland and Caffarel, Michel},
  journal={Physical Review Letters},
  volume={83},
  number={23},
  pages={4682},
  year={1999},
  publisher={APS},
  url={https://journals.aps.org/prl/abstract/10.1103/PhysRevLett.83.4682}
}

@inproceedings{cho2014learning,
  title={Learning phrase representations using RNN encoder--decoder for statistical machine translation},
  author={Cho, Kyunghyun and Van Merri{\"e}nboer, Bart and Gul{\c{c}}ehre, {\c{C}}a{\u{g}}lar and Bahdanau, Dzmitry and Bougares, Fethi and Schwenk, Holger and Bengio, Yoshua},
  booktitle={Proceedings of the 2014 conference on empirical methods in natural language processing (EMNLP)},
  pages={1724--1734},
  year={2014},
  url={https://aclanthology.org/D14-1179.pdf}
}

@article{hibat2020recurrent,
  title={Recurrent neural network wave functions},
  author={Hibat-Allah, Mohamed and Ganahl, Martin and Hayward, Lauren E and Melko, Roger G and Carrasquilla, Juan},
  journal={Physical Review Research},
  volume={2},
  number={2},
  pages={023358},
  year={2020},
  publisher={APS},
  url={https://journals.aps.org/prresearch/abstract/10.1103/PhysRevResearch.2.023358}
}

@article{metropolis1953equation,
  title={Equation of state calculations by fast computing machines},
  author={Metropolis, Nicholas and Rosenbluth, Arianna W and Rosenbluth, Marshall N and Teller, Augusta H and Teller, Edward},
  journal={The journal of chemical physics},
  volume={21},
  number={6},
  pages={1087--1092},
  year={1953},
  publisher={American Institute of Physics},
  url={https://www.osti.gov/biblio/4390578}
}

@article{hastings1970monte,
  title={Monte Carlo sampling methods using Markov chains and their applications},
  author={Hastings, W Keith},
  journal = {Biometrika},
  volume = {57},
  number = {1},
  pages = {97--109},
  year={1970},
  publisher={Oxford University Press},
  url = {https://doi.org/10.1093/biomet/57.1.97}
}

@article{viteritti2026approaching,
  title={Approaching the Thermodynamic Limit with Neural-Network Quantum States},
  author={Viteritti, Luciano Loris and Rende, Riccardo and Sachdev, Subir and Carleo, Giuseppe},
  journal={arXiv:2602.02665},
  year={2026},
  url={https://arxiv.org/abs/2602.02665}
}

@article{moss2025leveraging,
  title={Leveraging recurrence in neural network wavefunctions for large-scale simulations of Heisenberg antiferromagnets on the square lattice},
  author={Moss, M Schuyler and Wiersema, Roeland and Hibat-Allah, Mohamed and Carrasquilla, Juan and Melko, Roger G},
  journal={Physical Review B},
  volume={112},
  number={13},
  pages={134450},
  year={2025},
  publisher={APS},
  url={https://journals.aps.org/prb/abstract/10.1103/6ccd-wzhz}
}

@article{hornik1989multilayer,
  title={Multilayer feedforward networks are universal approximators},
  author={Hornik, Kurt and Stinchcombe, Maxwell and White, Halbert},
  journal={Neural networks},
  volume={2},
  number={5},
  pages={359--366},
  year={1989},
  publisher={Elsevier},
  url={https://www.sciencedirect.com/science/article/abs/pii/0893608089900208}
}

@article{lin2018resnet,
  title={Resnet with one-neuron hidden layers is a universal approximator},
  author={Lin, Hongzhou and Jegelka, Stefanie},
  journal={Advances in neural information processing systems},
  volume={31},
  year={2018},
  url={https://proceedings.neurips.cc/paper/2018/hash/03bfc1d4783966c69cc6aef8247e0103-Abstract.html}
}

@article{mann2025squeezing,
  title={Squeezing Classical Antiferromagnets into Quantum Spin Liquids via Global Cavity Fluctuations},
  author={Mann, Charlie-Ray and Oehlgrien, Mark A and Jaworowski, B{\'L} and Calaj{\u{A}}{\l}, Giuseppe and Marino, Jamir and Choi, Kyung S and Chang, Darrick E and others},
  journal={arXiv:2512.05630},
  year={2025},
  url={https://arxiv.org/abs/2512.05630v1}
}

@article{Chandrasekharan1997,
	title = {Quantum link models: {A} discrete approach to gauge theories},
	volume = {492},
	issn = {0550-3213},
	url = {https://www.sciencedirect.com/science/article/pii/S0550321397800417},
	doi = {https://doi.org/10.1016/S0550-3213(97)80041-7},
	number = {1},
	journal = {Nuclear Physics B},
	author = {Chandrasekharan, S. and Wiese, U.-J.},
	year = {1997},
	pages = {455--471},
}

@article{jvmc2022,
	title = {{jVMC}: {Versatile} and performant variational {Monte} {Carlo} leveraging automated differentiation and {GPU} acceleration},
	issn = {2949-804X},
	shorttitle = {{jVMC}},
	url = {https://scipost.org/10.21468/SciPostPhysCodeb.2},
	doi = {10.21468/SciPostPhysCodeb.2},
	language = {en},
	urldate = {2024-07-24},
	journal = {SciPost Physics Codebases},
	author = {Schmitt, Markus and Reh, Moritz},
	month = aug,
	year = {2022},
	pages = {002}
}

@article{rende2026transformer,
  title={Transformer Neural-Network Quantum States for lattice models of spins and fermions: Application to the Ancilla Layer Model},
  author={Rende, Riccardo and Nikolaenko, Alexander and Viteritti, Luciano Loris and Sachdev, Subir and Zhang, Ya-Hui},
  journal={arXiv:2603.02316},
  year={2026},
  url={https://arxiv.org/abs/2603.02316}
}

@article{chen2025neural,
  title={Neural network-augmented pfaffian wave-functions for scalable simulations of interacting fermions},
  author={Chen, Ao and Wan, Zhou-Quan and Sengupta, Anirvan and Georges, Antoine and Roth, Christopher},
  journal={arXiv preprint arXiv:2507.10705},
  year={2025},
  url={https://arxiv.org/abs/2507.10705}
}

@article{schneider2024dipolar,
  title={Dipolar ordering transitions in many-body quantum optics: Analytical diagrammatic approach to equilibrium quantum spins},
  author={Schneider, Benedikt and Burkard, Ruben and Olmos, Beatriz and Lesanovsky, Igor and Sbierski, Bj{\"o}rn},
  journal={arXiv preprint arXiv:2407.18156},
  year={2024},
  url={https://arxiv.org/abs/2407.18156}
}

@article{zhang2014quantum,
  title={Quantum phases in circuit QED with a superconducting qubit array},
  author={Zhang, Yuanwei and Yu, Lixian and Liang, J-Q and Chen, Gang and Jia, Suotang and Nori, Franco},
  journal={Scientific reports},
  volume={4},
  number={1},
  pages={4083},
  year={2014},
  publisher={Nature Publishing Group UK London},
  url={https://www.nature.com/articles/srep04083}
}

@article{lagnese2026neural,
  title={Neural network modeling of many-body super-and sub-radiant dynamics},
  author={Lagnese, Gianluca and Brunner, Laurin and Rossi, Lorenzo and Chang, Darrick and Schmitt, Markus and Lenar{\v{c}}i{\v{c}}, Zala},
  journal={arXiv preprint arXiv:2605.04640},
  year={2026},
  url={https://arxiv.org/pdf/2605.04640}
}

@misc{zenodo,
  author={Salmeron, Noe and Bukov, Marin and Schmitt, Markus},
  title={Data, code and models},
  year={2026},
  howpublished={\url{https://doi.org/10.5281/zenodo.20630702}}
}

\appendix

\section{Mean-Field theory}
\label{app:MF coherent}

We obtain the mean-field estimates by writing the wavefunction as a product state with three parameters $\{\alpha, \beta_e, \beta_o\}$:
\begin{widetext}
\begin{equation}
|\psi_{\text{MF}}\rangle=|\alpha\rangle\otimes\left(\cos{\beta_e}|0\rangle+\sin{\beta_e}|1\rangle\right)\otimes\left(\cos{\beta_o}|0\rangle+\sin{\beta_o}|1\rangle\right)\otimes\cdots
\end{equation}
\end{widetext}
The state of the cavity is a coherent state with parameter $\alpha$, and we treat the spins separately on the even and odd sublattices, with $\beta_e$ and $\beta_o$ respectively, to accommodate the antiferromagnetic ordering. For a better comparison with our neural network ansatz, we take all parameters to be real. The energy density is then given by
\begin{widetext}
\begin{equation}
E=\left\langle\psi_{\text{MF}}\left|\frac HN\right|\psi_{\text{MF}}\right\rangle=-\frac{\Delta}{2} \left(\cos ^2\beta_e+\cos ^2\beta_o-1\right)+V \cos ^2\beta_e\cos ^2\beta_o+\frac{g\alpha}{\sqrt N} (\sin (2 \beta_e)+\sin (2 \beta_o))+\frac{\alpha^2\omega_0}{N}
\end{equation}
\end{widetext}
This expression has to be minimized with respect to the parameters $\{\alpha, \beta_e, \beta_o\}$ to find the mean-field ground state. It can be made easier by explicitly computing $\alpha$ in terms of $\beta_e$ and $\beta_o$:
\begin{equation}
\frac{\partial E}{\partial \alpha}=0\implies\alpha=-\frac{g\sqrt N}{2\omega_0}(\sin (2 \beta_e)+\sin (2 \beta_o)).
\end{equation}
Leading to the following expression for the energy density:
\begin{widetext}
\begin{equation}
E=-\frac{\Delta}{2} \left(\cos ^2\beta_e+\cos ^2\beta_o-1\right)+V \cos ^2\beta_e\cos ^2\beta_o-\frac{g^2}{4\omega_0} (\sin (2 \beta_e)+\sin (2 \beta_o))^2
\end{equation}
\end{widetext}
The minimization with respect to $\beta_e$ and $\beta_o$ can then be carried out numerically to obtain the mean-field phase diagram in Fig~\ref{fig:fig1}(c).

It is also possible to obtain analytical expressions for the phase boundaries, which we detail in the rest of this section.

\paragraph{FP-SR phase boundary}
For $V/\Delta\ll1$ and $g/\Delta\ll1$, the external magnetic field dominates, and the ground state is fully polarized (FP) with $(\beta_e,\beta_o)=(0,0)$. From the numerical minimization, we know that the FP phase is surrounded by the superradiant (SR) phase, in which the spins are tilted toward the x axis at the same angle $\phi$: $(\beta_e,\beta_o)=(\phi,\phi)$. To probe the transition, we consider a small deviation $\delta\ll1$ around the FP state by writing $(\beta_e,\beta_o)=(\delta,\delta)$. The energy becomes
\begin{equation}
E=-\frac{\Delta}{2} \left(2 \cos ^2(\delta )-1\right)+V \cos ^4(\delta)-\frac{g^2}{\omega_0}\sin ^2(2 \delta )
\end{equation}
From the numerical minimization, we know that the phase transition is second order. The FP state, therefore, is no longer a minimum of the energy whenever the second derivative evaluated at $\delta=0$ changes sign. This occurs for
\begin{equation}
V/\Delta=\frac{1}{2}-\frac{2(g/\Delta)^2}{\omega_0/\Delta},
\end{equation}
providing the analytical expression for the FP-SR phase boundary.

\paragraph{AFM-Mixed phase boundary}
For $V/\Delta\gg1$ and $V/\Delta\ll g/\Delta$, the spin-spin interaction dominates, and the ground state is antiferromagnetic (AFM) with $(\beta_e,\beta_o)=(0,\pi/2)$ or $(\beta_e,\beta_o)=(\pi/2,0)$. To determine the boundary of the AFM phase, we consider a small deviation around one of the AFM ground states $(\beta_e,\beta_o)=(\delta_e,\pi/2-\delta_o)$ where $\delta_e,\delta_o\ll1$. The energy then becomes
\begin{align}
E=&-\frac{\Delta}{2} \left(\cos ^2\delta_e+\sin ^2\delta_o-1\right)+V \cos ^2\delta_e\sin ^2\delta_o\nonumber\\&-\frac{g^2}{4\omega_0} (\sin (2 \delta_e)+\sin (2 \delta_o))^2.
\end{align}
The numerical minimization predicts a second-order phase transition, except at $V/\Delta=1$. These second-order phase transitions occur when the Hessian determinant evaluated at $(\delta_e,\delta_o)=(0,0)$ changes sign, signaling that the AFM state transitions from a minimum to a saddle point of the energy. This condition leads to the following equation for the phase boundary:
\begin{equation}
\label{eq:AFM-Mixed}
V/\Delta= \frac{\omega_0/\Delta}{2\left(\omega_0/\Delta - 2 (g/\Delta)^2 \right)}.
\end{equation}
When the Hessian determinant is positive, we verify that the AFM state is indeed a minimum of the energy by making sure that the diagonal elements of the Hessian matrix are positive. This is the case for
\begin{equation}
(g/\Delta)^2<\frac{\omega_0/\Delta}{2}
\end{equation}
which is always satisfied for $g/\Delta$ and $V/\Delta$ positive and satisfying Eq.~\eqref{eq:AFM-Mixed}.

At the multi-critical point $V/\Delta=1$, the Hessian determinant still changes sign, meaning that the AFM state transitions from being a minimum to a saddle point of the energy. Because the transition is first order, this indicates that, exactly at the multi-critical point, for $(g/\Delta,V/\Delta)=(\sqrt{\omega_0/4\Delta},1)$, the minimum of the energy is achieved over a continuous number of states. Indeed, all mean-field states of the form $(\beta_e,\beta_o)=(\phi,\pi/2-\phi)$ for $\phi\in[0,\pi/2]$ share the same energy at this special point. This allows the minimum to jump discontinuously from $(\beta_e,\beta_o)=(0,\pi/2)$ in the AFM phase to $(\beta_e,\beta_o)=(\pi/4,\pi/4)$ in the SR phase.

This effect is visible in the energy at the multi-critical point. It reads
\begin{equation}
    E=\frac{\Delta}{8} (\cos (2 (\beta_e-\beta_o))+3) \cos ^2(\beta_e+\beta_o)\geq0
\end{equation}
and is minimized to $E=0$ whenever $\beta_e+\beta_o=\pi/2$.

\paragraph{SR-Mixed phase boundary}
For $g/\Delta\gg1$, the ground state is superradiant with no AFM ordering; in the mean-field picture, the spins are all aligned at an angle $\phi$ from the z axis: $(\beta_e,\beta_o)=(\phi,\phi)$. To find the angle $\phi$, we plug it into the energy and find the minimum, yielding:
\begin{equation}
\phi=\cos^{-1}\left(\sqrt{\frac{4g^2+\Delta\omega_0}{8g^2+2V\omega_0}}\right).
\end{equation}
The transition to the mixed phase occurs when the unit cell translation symmetry breaks, leading to a different alignment of the spins on each sublattice. We can obtain the analytical expression of this phase boundary by considering a small deviation $\delta\ll1$ around the SR state that breaks the unit cell translation symmetry $(\beta_e,\beta_o)=(\phi+\delta,\phi-\delta)$. Contrary to the AFM-Mixed phase boundary, we do not need to consider a different variation on each sublattice to recover the correct phase boundary found via numerical minimization. The energy reads
\begin{align}
\label{eq:SR-Mixed energy}
E=&-\frac{\Delta}{2} \left(\cos ^2(\phi-\delta )+\cos ^2(\phi+\delta )-1\right)\nonumber\\&+V \cos ^2(\phi-\delta ) \cos ^2(\phi+\delta )\nonumber\\&-\frac{g^2}{4 \omega_0} (\sin (2 (\phi-\delta ))+\sin (2 (\phi+\delta )))^2.
\end{align}
Once again, we know from the numerical minimization that the phase transition is second order everywhere, except at $V/\Delta=1$. The phase boundary is therefore given by the points where the second derivative evaluated at $\delta=0$ changes sign. This yields the final phase boundary
\begin{widetext}
\begin{equation}
V/\Delta=\frac{16 (g/\Delta)^4+(\omega_0/\Delta)^2+\left(4 (g/\Delta)^2+(\omega_0/\Delta) \right) \sqrt{80 (g/\Delta)^4-8 (\omega_0/\Delta) (g/\Delta)^2+(\omega_0/\Delta) ^2}}{4 (\omega_0/\Delta)\left(2 (g/\Delta)^2+\omega_0/\Delta \right)}.
\end{equation}
\end{widetext}
At the multi-critical point, the second derivative still changes sign, leading to the same implications described in the AFM-Mixed phase boundary derivation.

There is an alternative mean-field approach detailed in Ref.~\cite{gelhausen2016quantum}, in which the Hamiltonian is first transformed to an effective spin-only Hamiltonian by integrating out the quadratic photon term.

\section{Optimization details}
\label{app:optimization}

We use the Stochastic Reconfiguration method to update the parameters $\theta_j$ of the neural network \cite{sorella2005wave, hackl2020geometry}. The update to the parameters $\delta \theta_j$ are computed by solving the linear equation
\begin{equation} 
    \label{eq:SR} \sum_{j} S_{ij} \delta \theta_j = -F_i, 
\end{equation}
where $S_{ij}$ is the quantum geometric tensor, and $F_i$ is the gradient of the energy with respect to the parameters. The components of these two quantities are given by
\begin{equation}
    \label{eq:Sij} S_{ij} = \real{ \mathbb{E}_{\pi} \left[ \Gamma_i^*(x) \Gamma_j(x) \right] - \mathbb{E}_{\pi} \left[ \Gamma_i^*(x) \right] \mathbb{E}_{\pi} \left[ \Gamma_j(x) \right] }, 
\end{equation}
and
\begin{equation} 
    \label{eq:gi} F_i = \real{ \mathbb{E}_{\pi} \left[ E_\text{loc}(\boldsymbol{x}) \Gamma_i^*(x) \right] - \mathbb{E}_{\pi} \left[ E_\text{loc}(x) \right] \mathbb{E}_{\pi} \left[ \Gamma_i^*(x) \right] }.
\end{equation}
Here, $\Gamma_i(x) = \partial_{\theta_i}\text{ln } \psi_{\boldsymbol\theta}(x)$ is the log-derivative of the wavefunction with respect to the parameter $\theta_i$. 

In practice, both $S_{ij}$ and $F_i$ are computed as expectation values over the Born distribution $\pi$ and can therefore be estimated using the same set of samples used for evaluating the energy.

The quantum geometric tensor $S_{ij}$ defined in Eq.~\eqref{eq:SR} has size $N_p\times N_p$, where $N_p$ is the number of parameters in the network. Inverting this tensor, therefore, scales as $\mathcal O(N_p^3)$. To accommodate large network sizes, we use the minSR trick \cite{chen2024empowering, rende2024simple}, which reduces the scaling to $\mathcal O(N_p)$. At each iteration, we use between 2048 and 3072 samples to estimate the parameter update.

For the simulations used to produce Fig.~\ref{fig:cuts}, we tune the hyperparameters individually to ensure stable optimization and reasonable convergence time. The timesteps range from $dt=0.002$ to $dt=0.03$. In general, larger values to the spin-photon coupling $g/\Delta$ required smaller timesteps for a stable optimization. We attribute this to the increased number of photons in the ground state for stronger coupling, making the optimization more challenging. In a similar fashion, the cutoff has to be increased as the spin-photon coupling grows. The complete list of hyperparameters used for each simulation in Fig.~\ref{fig:cuts} can be found in the \textit{simulation\_parameters.json} file in the Zenodo repository~\cite{zenodo}.

\begin{figure}[]
\includegraphics[width=0.5\textwidth]{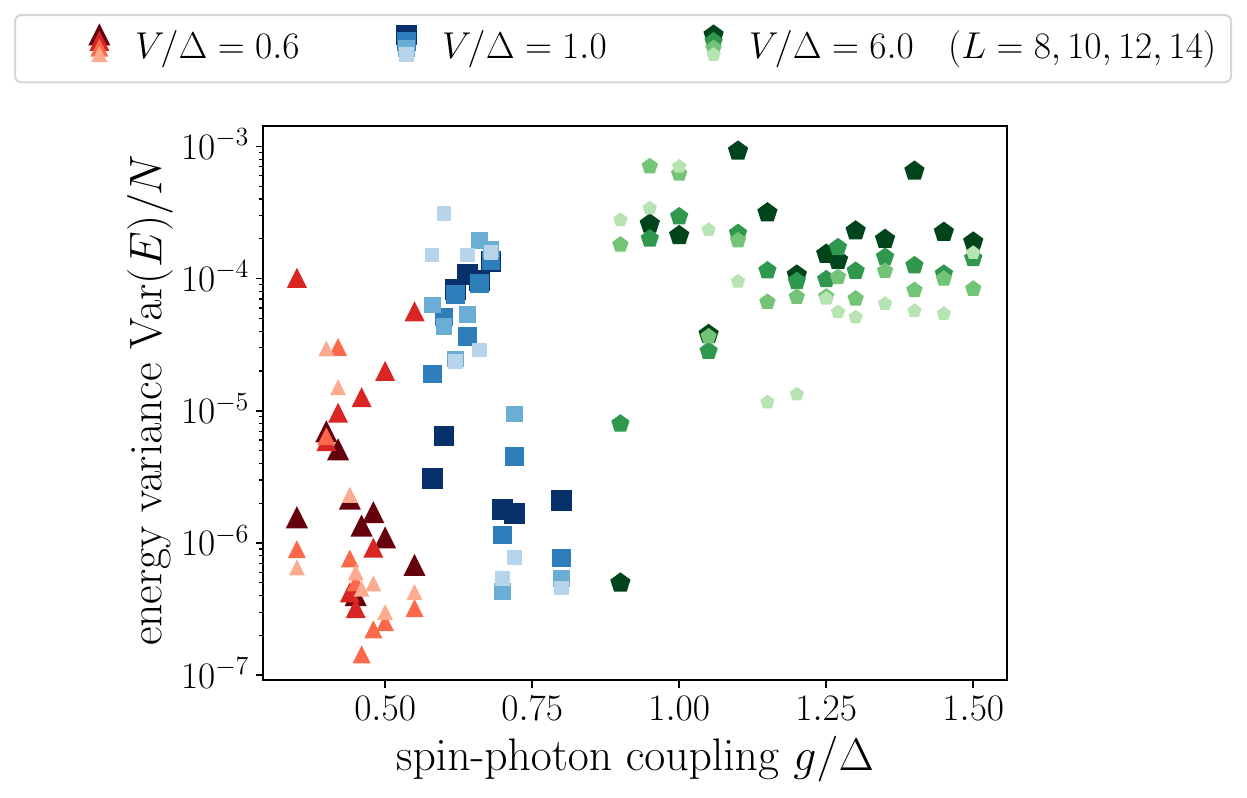}
\caption{\label{fig:app-variance} Energy variance per spin for each simulation in Fig.~\ref{fig:cuts}. The values are consistently below $10^{-3}$, indicating proper convergence to an eigenstate of the Hamiltonian.
}
\end{figure}

Thanks to the zero-variance property \cite{assaraf1999zero}, the energy variance can serve as an indicator for how close the variational state is to an eigenstate of the Hamiltonian. We plot in Fig.~\ref{fig:app-variance} the variance reached by each simulation used in Fig.~\ref{fig:cuts}. The values consistently lie below $10^{-3}$, indicating the proximity of an eigenstate.

\section{Neural network details}
\label{app:nn}

We use the neural network architecture described in Section~\ref{sec:architecture}, consisting of eight residual blocks, each containing two convolutional layers. The convolutional layers have either 8 or 16 channels. Increasing the number of residual blocks or channels did not lead to a significant improvement in the variational energy. The weights are drawn from the He-normal distribution \cite{he2015delving}, a truncated normal distribution with mean 0 and standard deviation $\sqrt{2/n_{\mathrm{in}}}$, where $n_{\mathrm{in}}$ is the number of inputs to the current layer. The biases in the convolutional and fully connected layers, on the other hand, are initialized to 0. Between convolutional layers, we apply the GELU activation function \cite{hendrycks2016gaussian}.

Each convolutional layer uses a $3\times 3$ kernel with stride 1 and circular padding. These choices ensure that the convolutions are equivariant under a translation of the input. After the final convolutional layer, the data goes through a global average pooling layer where each feature map is averaged to one number. This global pooling operation ensures that the overall network remains translationally invariant.

We initially experimented with different neural network architectures, but they could not match the variational energies of the multi-head ResNet or suffered from numerical instabilities. We first tried using a recurrent neural network (RNN) with Gated Recurrent Units (GRU) \cite{cho2014learning}. For a configuration $\ket{n,\boldsymbol\sigma}$, the corresponding wavefunction coefficient is the square root of the Born distribution noted $P(n,\boldsymbol\theta)$: $\psi(n,\boldsymbol\theta)=\sqrt{P(n,\boldsymbol\theta)}$~ \cite{hibat2020recurrent}. The RNN we used computes this value as $P(n,\boldsymbol\theta)=P(n)\prod_{j=1}^N P(\sigma_j|n,\sigma_1,\dots\sigma_{j-1})$ where the photon number is the first input of the network, followed by the spins. Because of instabilities and poor variational energies, we then turned to a CNN-based ansatz, better suited to translationally invariant two-dimensional problems. We encoded the photon degree of freedom in the input by multiplying its value with the spins: $n\times(\sigma_1, \sigma_2, \dots, \sigma_N)$ where $\sigma_j\in{-1,+1}$. This architecture suffered from the same issues as the RNN, prompting us to develop the multi-head ResNet presented in this paper.

\end{document}